\documentclass[%
 reprint,
 amsmath,amssymb,
 aps,
 prapplied,
]{revtex4-2}

\usepackage{graphicx}
\usepackage{dcolumn}
\usepackage{bm}

\usepackage[version=4]{mhchem}

\usepackage[shortcuts]{extdash}
\usepackage{upgreek}
\usepackage{color}

\usepackage{xfrac}    

\usepackage{xargs} 

\newcommandx{\carsten}[2][1=]{}
\newcommandx{\tobias}[2][1=]{}

\begin{document}
\newcommand{\IGN}{\ce{In_xGa_{1-x}N}}
\newcommand{\xray}{\mbox{X-ray}}
\newcommand{\xrays}{\mbox{X-rays}}
\newcommand{\Fig}[1]{Fig.~\ref{#1}}
\newcommand{\unit}[1]{ \,\mathrm{#1} } %
\newcommand{\angstrom}{ \,\textup{\AA} } %
\newcommand{\Eq}[1]{Eq.~(\ref{#1})}
\newcommand{\Eqs}[1]{Eqs.~(\ref{#1})}
\newcommand{\Tab}[1]{Table~\ref{#1}}

\preprint{APS/123-QED}

\title{Nanoscale mapping of the full strain tensor, rotation and composition in partially relaxed \ce{In_xGa_{1-x}N} layers by scanning X-ray diffraction microscopy}


\author{Carsten Richter}
\email{carsten.richter@ikz-berlin.de}
\affiliation{Leibniz-Institut f\"ur Kristallz\"uchtung (IKZ), Max-Born-Str. 2, 12489 Berlin, Germany}
\affiliation{European Synchrotron Radiation Facility, BP 220, 38043 Grenoble Cedex, France}

\author{Vladimir M. Kaganer}
\affiliation{Paul-Drude-Institut f\"ur Festk\"orperelektronik (PDI), Hausvogteiplatz 5--7, 10117 Berlin, Germany}

\author{Armelle Even}%
\altaffiliation[Now at:\ ]{OSRAM Opto Semiconductors GmbH, 93055 Regensburg, Germany}
\affiliation{University of Grenoble-Alpes, CEA, LETI, Minatec Campus, F-38054 Grenoble, France}

\author{Am\'elie Dussaigne}
\affiliation{University of Grenoble-Alpes, CEA, LETI, Minatec Campus, F-38054 Grenoble, France}

\author{Pierre Ferret}
\affiliation{University of Grenoble-Alpes, CEA, LETI, Minatec Campus, F-38054 Grenoble, France}

\author{Fr\'ed\'eric Barbier}
\affiliation{University of Grenoble-Alpes, CEA, LETI, Minatec Campus, F-38054 Grenoble, France}

\author{Yves-Matthieu Le Vaillant}
\affiliation{Nelumbo Digital SAS,
 143 rue du Brocey, 38920 Crolles, France  }

\author{Tobias U. Sch\"ulli}
\affiliation{European Synchrotron Radiation Facility, BP 220, 38043 Grenoble Cedex, France}

\date{\today}

\begin{abstract}
Strain and composition play a fundamental role in semiconductor physics, since they are means to tune the electronic and optical properties of a material and hence develop new devices. Today it is still a challenge to measure strain in epitaxial systems in a non-destructive manner which becomes especially important in strain-engineered devices that often are subjected to intense stress. In this work, we demonstrate a microscopic mapping of the full tensors of strain and lattice orientation by means of scanning \xray{} diffraction microscopy. We develope a formalism to extract all components of strain and orientation from a set of scanning diffraction measurements and apply the technique to a patterned \IGN{} double layer to study strain relaxation and indium incorporation phenomena. The contributions due to varying indium content and threading dislocations are separated and analyzed.
\end{abstract}

\maketitle


\section{Introduction}

Epitaxial thin films are the basis for most of modern semiconductor (opto-)electronic devices, such as light emitting diodes (LEDs), transistors, integrated circuits etc. \cite{Sun2010}. Lattice strain caused by a mismatch of lattice constants or thermal expansion coefficients between film and substrate is an important consequence of the growth, but also an opportunity to influence the electronic and optical properties of the material. Next to varying the alloy composition, strain engineering therefore is an established route to tune these properties. 

Today, while transmission electron microscopy yields atomic resolution maps of lattice parameters \cite{Bierwolf1993, Schulz2014}, a nondestructive microscopic characterization of lattice strain with high accuracy is still not done routinely. Raman spectroscopy is used to study the impact of strain or alloy composition on phonon modes \cite{Wagner1999, Capellini2013}, allowing to formulate empirical models to indirectly relate strain to the measured Raman shift \cite{Gassenq2016}. However, these are usually restricted to a certain component of the strain tensor and are limited to a range of known materials. Recent developments of scanning electron microscopy (SEM) techniques provide microscopic access to lattice deformations. High resolution electron backscattered diffraction (HR-EBSD) patterns analyzed by cross-correlation lead to an improved sensitivity to relative strain and rotation down to $10^{-4}$, which has been used to characterize local dislocation densities \cite{Wilkinson2010, Wilkinson2011}. Electron channeling contrast imaging (ECCI) produces qualitative maps of lattice deformations with a high spatial resolution \cite{NareshKumar2012} that display individual dislocations and allow to determine their type even at high dislocation densities of $10^{10}\,\mathrm{cm}^{-2}$. A combination of these two techniques was used to determine the densities of screw, edge and mixed dislocations in epitaxial InAlN films \cite{VilaltaClemente2017}.

On the other hand, synchrotron based \xray{} diffraction microscopy techniques provide high lattice sensitivity and experienced an intensive development based on improved optics \cite{Schlli2018}. Focusing the beam now enables scanning \xray{} diffraction microscopy (SXDM) measurements with a resolution down to tens of nanometers \cite{Chahine2014,Chayanun2019}. The use of a polychromatic (`white`) beam leads to the simultaneous excitation of several Bragg reflections and the resulting Laue patterns provide information about the full strain tensor \cite{Robach2013, Tardif2016}. Furthermore, in the case of nano- and micro-particles, the coherence of the synchrotron radiation and recent developments in phase retrieval allow for the lens-less imaging of selected components of strain in 3d with a resolution below the beam size \cite{Pfeifer2006,Hofmann2017, Hofmann2020}. 

In this work, we focus on the development of SXDM to extract the full strain tensor, lattice orientation and alloy composition in a semiconductor heterostructure. Compared to other techniques, SXDM stands out for being non-destructive, providing high lattice sensitivity below $10^{-5}$ of rotation and strain \cite{Chahine2014}, access to buried layers and the compatibility with complex sample environments \cite{Richard2020}. Today, the spatial resolution at state-of-the-art beamlines is in the several $10\,\mathrm{nm}$ range and SXDM gives direct, model-free information about the lattice parameters.
Although the technique has already been applied to several material systems and devices to study strain and composition \cite{Zoellner2015, Zoellner2019}, former results were based on limited data and relied on the average crystallographic symmetry. However, on the microscopic scale, no symmetry of the unit cell can be presumed because of an anisotropic local stress. This implies that all lattice parameters need to be refined in order to correctly decouple isotropic expansion of the lattice due to alloying and the anisotropic elastic strain as response to (local) stress. Hence, the knowledge of the full strain tensor is required which also allows to infer local dislocation densities \cite{VilaltaClemente2017}. In \xray{} diffraction (XRD), this means that the reciprocal space position of at least three non-coplanar Bragg reflections needs to be known for each real space position, as is demonstrated below.

We used the SXDM technique to study strain relaxation in an \IGN{} (InGaN) heterostructure that serves as a template for the growth of multiple quantum well (MQW) structures. III-Nitride semiconductors (GaN, InN, AlN and their alloys) received huge attention and extensive development for their applications in electronics \cite{mi2017iii,Gil2013}, particularly as light emitting diodes (LEDs) \cite{Akasaki1997, nakamura2013blue} and power electronics \cite{Baliga_2013}.
The interest in the InGaN alloy is based on the potential of a direct electroluminescence at any wavelength of the visible spectrum \cite{Wu:2009} by tuning the In concentration. White light emission is nowadays realized by indirect color conversion using phosphors excited with a short-wavelength nitride LED \cite{Narukawa2010}. However, for future micro-displays with pixel size below $10\unit{\upmu m}$, a monolithic integration becomes advantageous. InGaN based MQWs are the most promising candidates for this purpose but they still suffer from reduced efficiency at longer emission wavelengths due to limitations in the currently achievable In-concentration of $<30\,\%$ \cite{Pasayat2020}.

One approach for growing high quality InGaN films with higher In content is to reduce the mismatch strain which is a result of pseudomorphic growth on GaN substrates and the larger covalent radius of In compared to Ga. It has been shown for such strained films that a preference of In to occupy fourfold coordinated surface sites poses a limit to the indium concentration to 0.25 \cite{Lymperakis2018}. The large mismatch also results in mechanical stress at the interface and may result in additional defects like dislocations or V-pits as the film thickness increases \cite{Dussaigne2020}.
An obvious way to reduce the mismatch is to grow InGaN films on strain-relaxed InGaN buffer layers that have an in-plane lattice parameter close to the one of the functional InGaN film. An example for such a virtual substrate is InGaN on sapphire (InGaNOS) from Soitec \cite{Even2017, Dussaigne2020b}, which is transferred from an initial InGaN/GaN donor structure using the Smart Cut\textsuperscript{TM} technique \cite{Tauzin2005}. After transfer, the InGaNOS seed layers are structured and annealed to facilitate strain relaxation and hence provide optimized in-plane lattice parameters for subsequent regrowth. 

Here we demonstrate the capabilities of SXDM to map of the full tensors of lattice strain and rotation, allowing to investigate the relaxation mechanisms in both layers of an InGaN/InGaNOS heterostructure. The results are discussed in the context of partial strain relaxation, variations of alloy composition and strain fields around threading dislocations.

\section{Experiment}
\subsection{Sample structure}

\begin{figure}
    \includegraphics[width=0.9\columnwidth]{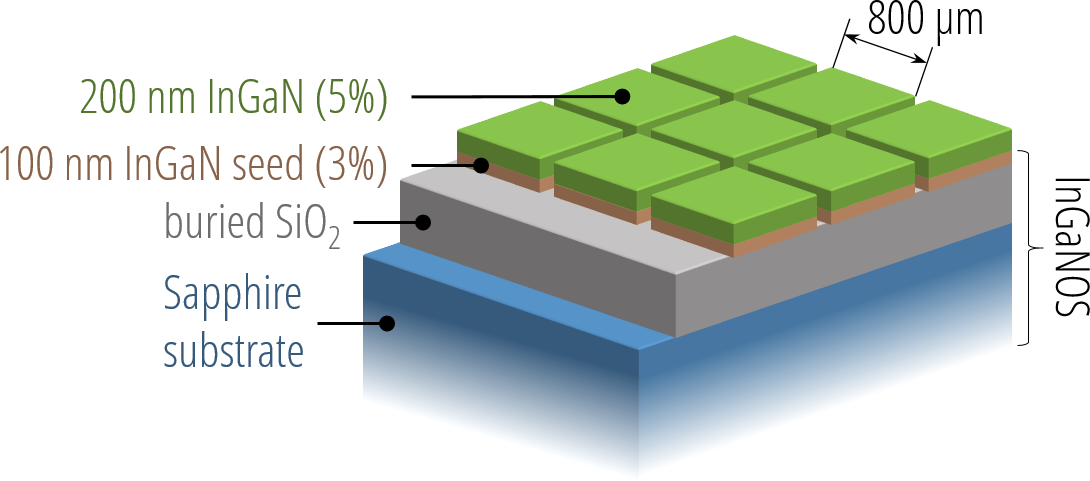}
    \caption{\label{fig:stack} Sketch of the sample (not drawn to scale). An InGaN seed layer with nominal In content of $3\,\%$ (orange) has been bonded onto \ce{SiO_2} coated Sapphire. After patterning and heat treatment, a layer with the In content of $5\,\%$ (green) was epitaxially grown on top of the thus obtained InGaNOS virtual substrate.}
\end{figure}

Our sample is based on an InGaNOS pseudo substrate from Soitec with nominal in-plane lattice parameter of
$a=3.190\unit{\textup{\AA}}$
corresponding to an In-content of $3\,\%$, see Fig.\ \ref{fig:stack}. The (0001) oriented InGaNOS was obtained through growth by metal-organic vapor phase epitaxy (MOVPE) of an InGaN seed layer on GaN and subsequent bonding of the this layer onto a SiO$_2$ coated sapphire substrate. 
For strain relaxation, mesa structures with a side length of $800\unit{\upmu m}$ have been patterned by photolithography and dry etching, followed by several annealing steps. After the transfer and patterning, the InGaNOS seed layer thickness amounts to $\approx 100\unit{nm}$. Almost full relaxation of such mesa structures has been reported and they have successfully been used as substrate for MQWs emitting in nearly the full visible range \cite{Even2017, Dussaigne2020}.
A second \IGN{} layer of higher (nominally $5\,\%$) In-content and a thickness of  $\approx 200\unit{nm}$ has been grown on top of the InGaNOS pseudo substrate by MOVPE. The relaxation of the InGaNOS reduces the lattice mismatch of the two layers and this way enables higher In uptake during epitaxy. Due to the patterning of the substrate, enhanced relaxation is to be expected at the edges of each pad. In this region, also a higher density of V-pits can be seen in scanning electron microscopy (SEM) images (see \Fig{fig:intensity}(a)).

\subsection{Scanning X\=/ray diffraction microscopy}

\begin{figure}
\includegraphics[width=0.8\columnwidth]{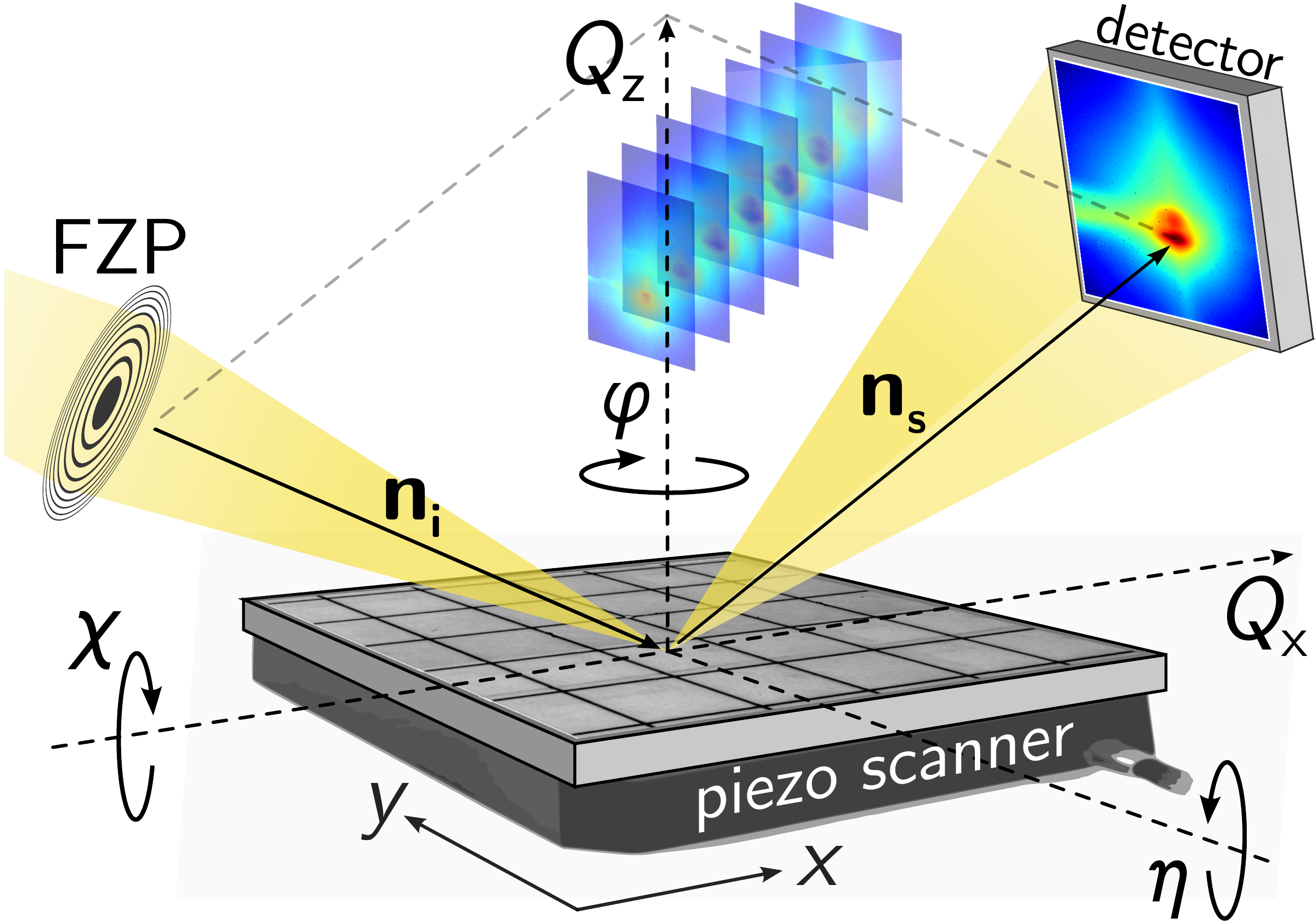}
\caption{\label{fig:setup} Illustration of the SXDM setup with definition of the coordinate system. The beam is focused (e.g. by a Fresnel Zone Plate (FZP)) onto the sample. While the sample is scanned through the beam, the detector continuously records images of diffracted intensity. Each frame corresponds to a 2d slice of the 3d reciprocal space $\mathbf Q$ which is defined by the direction of incident and scattered beam and the X\=/ray wavelength $\lambda$: $\mathbf Q = 2\pi/\lambda \left( \mathbf n_s - \mathbf n_i \right)$. By rocking about the $\eta$-axis, a 3d volume of reciprocal space is probed.}
\end{figure}

SXDM measurements have been carried out at beamline ID01 of the ESRF using a focused X\=/ray beam with sub-micron spot size and a fast, piezo-driven scanning stage (see \Fig{fig:setup}). Details of the setup have been described by Chahine et al. \cite{Chahine2014}.
Continuous scans were performed unidirectionally by driving the piezo stage at constant speed during synchronized readout of the area detector (MaxiPix 4) \cite{Ponchut2007} at a frame-rate of $100\unit{s^{-1}}$. The detector samples a 2d region of reciprocal (angular) space defined by its pixel size ($55\unit{\upmu m}$), the number of pixels ($516\times 516$), its distance to the sample ($\approx 670 \unit{mm}$) and the \xray{} energy (see below). By subsequently changing the angle of beam incidence $\eta$, a 3d reciprocal space map (RSM) $I(i, j, \eta)$ is probed. Here ($i,j$) are the row and column of the detector, respectively. Based on a calibration of the detector position and orientation, we convert these data to Cartesian coordinates of reciprocal space $I(Q_x, Q_y, Q_z)$. This is done for all points ($x,y$) on the sample surface resulting in a 5d intensity distribution $I(x, y, Q_x, Q_y, Q_z)$. Before this conversion, a drift correction is usually needed in order to compensate the parasitic motion of the sample on changing the incidence angle which is due to limitations in the alignment and the rigidity of the setup. The drift is typically in the range below $1 \mathrm{\upmu m}/ \mathrm{deg}$. The criterion we used to assess the amount of drift was to minimize the spatial variation of integrated intensity which is increased by any drift of the sample. The correction is then performed by sub-pixel shifting the maps $I_{i,j,\eta}(x,y)$ for each set of $i,j,\eta$.

\begin{figure}
    \begin{flushright}
        \includegraphics[width=.98\columnwidth]{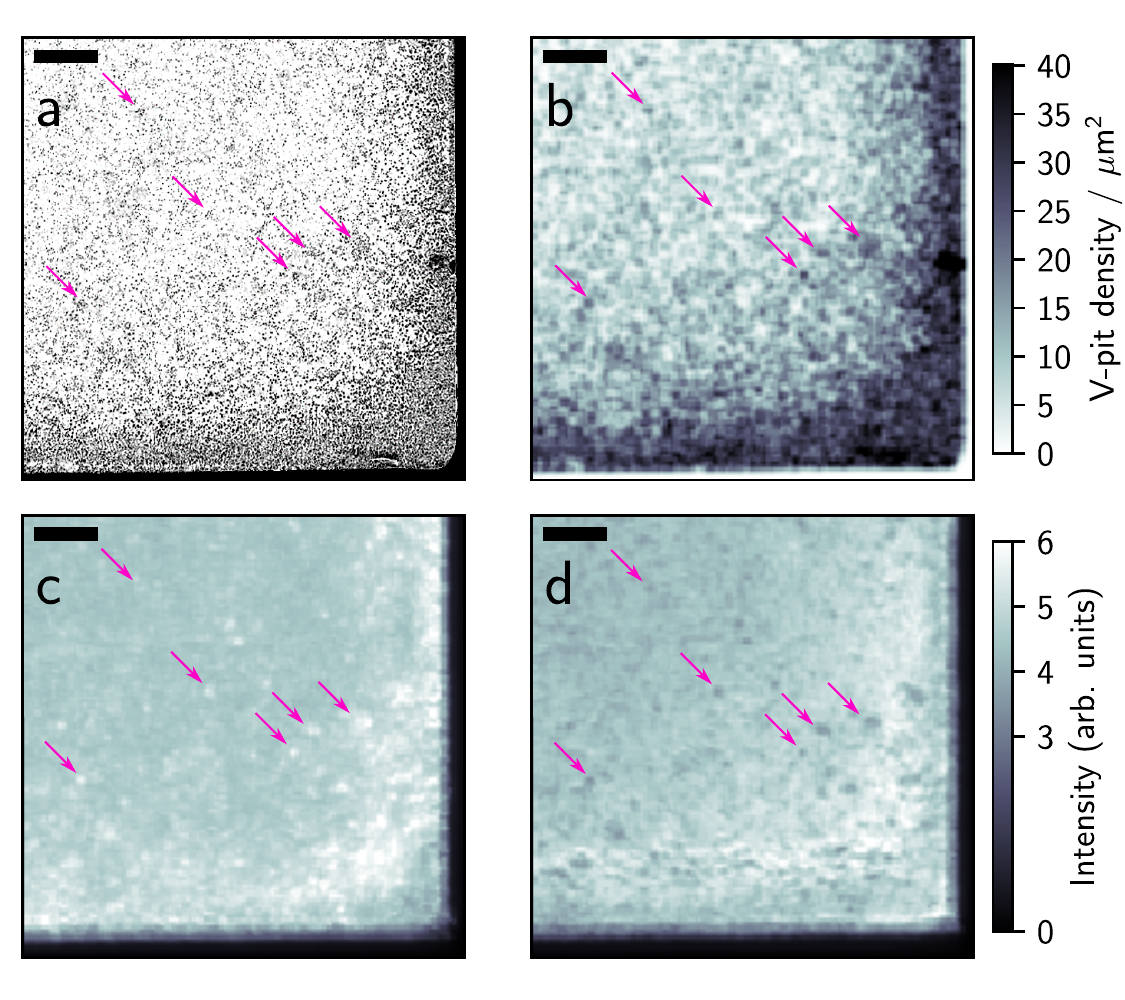}
        \end{flushright}
\caption{\label{fig:intensity} Scanning electron micrograph (SEM) of the studied corner region (a) compared to integrated intensity from SXDM of the InGaNOS seed layer (c) and the overgrown InGaN layer (d). The dark dots in (a) are V-pits in the InGaN film. Sub figure (b) shows the V-pit density which has been approximated based on the SEM image. One can see from (b) and (d) that clusters of V-pits result in a minimum of diffracted intensity. The arrows point at some clusters of V-pits that are seen in all datasets. The scale bars correspond to $5\,\mathrm{\upmu m}$.}
\end{figure}

The surface normal of the (0001) oriented InGaN/InGaNOS sample has been chosen as $Q_z$ direction, which corresponds to the vertical direction when all diffractometer angles are zero. $Q_x$ was chosen to be along the in-plane direction $[10\bar 10]$ (see coordinate system in \Fig{fig:setup}). 
We acquired 5d SXDM data sets in coplanar geometry for three different Bragg reflections, the symmetric $0004$ reflection and the asymmetric pair $10\bar 13$ and $0\bar 113$, to provide sensitivity to all lattice parameters. For each of them, we measured the same sample area of $40 \times 40\,\mathrm{\upmu m}^2$ at a corner of a partially relaxed mesa structure with a resolution of $150\times 150$ points resulting in a step size of $267\,\mathrm{nm}$ in both directions. \Fig{fig:intensity}(c),(d) shows the average intensity (integrated over reciprocal space $Q_x, Q_y, Q_z$) obtained for this region next to the corresponding SEM image.

The diffracted intensity and the SEM image show similar features. For easier comparison, the SEM image (\Fig{fig:intensity}(a)) has been resampled (\Fig{fig:intensity}(b)) to match the resolution of the \xray{} measurement (\Fig{fig:intensity}(c,d)). This way, \Fig{fig:intensity}(b) approximately illustrates the density of V-pits as gray-scale. Comparing this to the \xray{} intensities, one can see that a high V-pit density results in a reduced diffraction intensity for the top layer (\Fig{fig:intensity}(d)). This is expected since a V-pit means a loss of diffraction volume. However, an increase of intensity from the InGaNOS seed layer is observed. This may be explained by a strong local reduction of indium concentration due to the V-pit which causes a shift of the Bragg peak from the position of the In-rich top layer towards the seed layer with lower In-content. Therefore a change of the corresponding intensity ratios is observed. Below we will quantify the changes of In-content based on a combination of multiple SXDM measurements.

SXDM measurements of symmetric and asymmetric reflections have been carried out during two different experimental sessions and therefore under slightly different beam conditions. The symmetric reflection has been measured using \xrays{} with energy of $8\,\mathrm{keV}$ focused by a Fresnel zone plate (FZP) down to $(h,v)\approx(130,90)\,\mathrm{nm}$ in horizontal and vertical direction, respectively. On the other hand, for the asymmetric reflection the energy has been tuned to $7\,\mathrm{keV}$ providing the possibility to have almost normal incidence of \xrays{} and, hence, a smaller footprint and higher spatial resolution. For the latter case, a Kirkpatrick-Baez mirror system was used for focusing down to $(h,v)\approx(150,200)\,\mathrm{nm}$. Overall, the beam footprint is $200\,\mathrm{nm}$ or below which is smaller than the scan step size.

\section{Results and Discussion}
\subsection{Reciprocal space maps}

%
\begin{figure}
\includegraphics[width=\columnwidth]{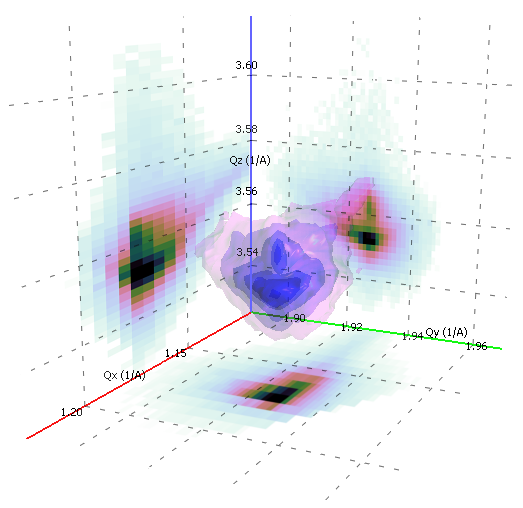}
\caption{\label{fig:qavg} Intensity distribution in 3d reciprocal space map (RSM) in vicinity of the $0\bar 113$ reflection for a single spot on the sample and projections along the Cartesian axes $Q_{z} \parallel 0001$ (blue), $Q_{x}\parallel 10\bar 10$ (red) and $Q_{y}$ (green). The spacing of intensity isosurfaces corresponds to a factor of three in intensity.}
\end{figure}

\Fig{fig:qavg} shows one of the collected 3d RSMs and its projections along the principle Cartesian axes as example for the $0\bar 113$ reflection. The data shows some typical features expressed in many parts of the sample. Two distinct maxima at different $Q_{z}$ positions correspond to the two pseudomorphic InGaN layers: the InGaNOS virtual substrate (at higher $Q_z$) and the top InGaN layer (stronger peak at lower $Q_z$). Streaks in the angular directions (rotation about $Q_{x}$ and $Q_{y}$) are due to local lattice tilt. The main goal of the data reduction is to extract the center position of both peaks from each of the (in total 67500) 3d RSMs. 



The local lattice deformations are studied in terms of the orientation and strain state of the crystallographic unit cells. The intensity distribution in reciprocal space is related to the strain and tilt distribution of the unit cells averaged over the probed volume. In our case, it was defined by the beam footprint and the layer thickness and is on the order of $\approx\! 150\!\times\! 200\! \times\! 150 \unit{nm^3}$. Furthermore, we only analyze the average lattice properties in such volume element, which is derived from the peak position (e.\,g. center of mass of intensity) in reciprocal space $\mathbf{Q}_c = (\overline{Q_x}, \overline{Q_y}, \overline{Q_z})$. Representing $\mathbf{Q}_c$ in spherical coordinates, one directly obtains the interplanar spacing $d$ of the diffracting lattice planes from the radial component $d = 2\pi / |\mathbf{Q}_c|$, whereas the angular components contain contributions from both tilt and strain. 
Due to the similar lattice parameters of the two investigated InGaN layers, we had to fit two strongly overlapping peaks (see \Fig{fig:qavg}) to determine positions of both peaks. For each RSM, we performed a double-peak fit on radial projections ($I(x, y, |\mathbf Q|$), since these provided the clearest peak separation. For both of the two resulting peak positions along $|\mathbf Q|$, we calculated the centers of mass for the two other (angular) projections and converted the results back to Cartesian coordinates. Thus, we reduced the 5d dataset $I(x, y, Q_x, Q_y, Q_z)$ to vector fields of the type $\mathbf Q_c (x, y)$ for each of the three measured reflections.
In order to combine the $\mathbf{Q}_c$ data of the three non-coplanar Bragg reflections (which allows to disentangle tilt and strain) one needs to match the surface coordinates $(x,y)$ of all data sets. This was achieved by correlating the angular components of $\mathbf Q_c$ that carry a contribution of the local lattice tilt for all reflections.

\subsection{Derivation of local lattice parameters and orientation}

The general equation relating the measured momentum transfer $\mathbf Q$ of a certain reflection $hkl$ with the crystal orientation and lattice parameters is \cite{Lohmeier1993}
\begin{equation}
\label{eq:q_UB}
\begin{pmatrix}Q_x \\ Q_y \\Q_z \end{pmatrix} = \mathbf U(\varphi,\chi,\eta)\cdot \mathbf B(a,b,c,\alpha,\beta,\gamma) \cdot \begin{pmatrix}h  \\ k \\l \end{pmatrix}.
\end{equation}
Assuming that the crystallographic symmetry is locally fully broken by strain (space group \textit{1}), the matrix $\mathbf B$ depends on all six lattice parameters ($a,b,c,\alpha,\beta,\gamma$) while the matrix $\mathbf U$ takes account of the local orientation of the probed crystalline volume, e.\,g. in terms rotation about the three Cartesian axes ($\chi, \eta, \varphi$, see \Fig{fig:setup}). Thus, the system of equations (\ref{eq:q_UB}) containing 9 unknowns becomes determined with the data collected for at least three non-coplanar reflections. 

In practice, relative variations in the components of $\mathbf{Q}_c$ can be determined very precisely. However, the accuracy of absolute values is limited and leads to offsets in the resulting lattice parameters. This is mainly due to the sphere of confusion of the diffractometer, its limited stiffness, but also the fact that we remounted the sample for one of the measurements. It is critical to correct the average offsets of $\mathbf Q_c$ by using a reference, for example, based on reflections of a known substrate or by comparing the sample average of $\mathbf{Q}_c$ to lab-based, integrating XRD measurements. We used the latter approach and also assumed that, on average, the hexagonal symmetry of the lattice holds. With these additional conditions, we correct the $(\overline{Q_x}, \overline{Q_y}, \overline{Q_z})$ values and determine the average orientation matrix $\mathbf{U}$. Subsequently, local variations of orientation and lattice parameters can be determined by solving \Eqs{eq:q_UB} for each point on the sample and for both InGaN layers. Local orientation is described by applying an additional rotation matrix $\Delta \mathbf U (\Delta \varphi, \Delta \chi, \Delta \eta)$ whereas lattice strain is treated by refining the components of $\mathbf B$.

\begin{figure}
    \includegraphics[width=\columnwidth]{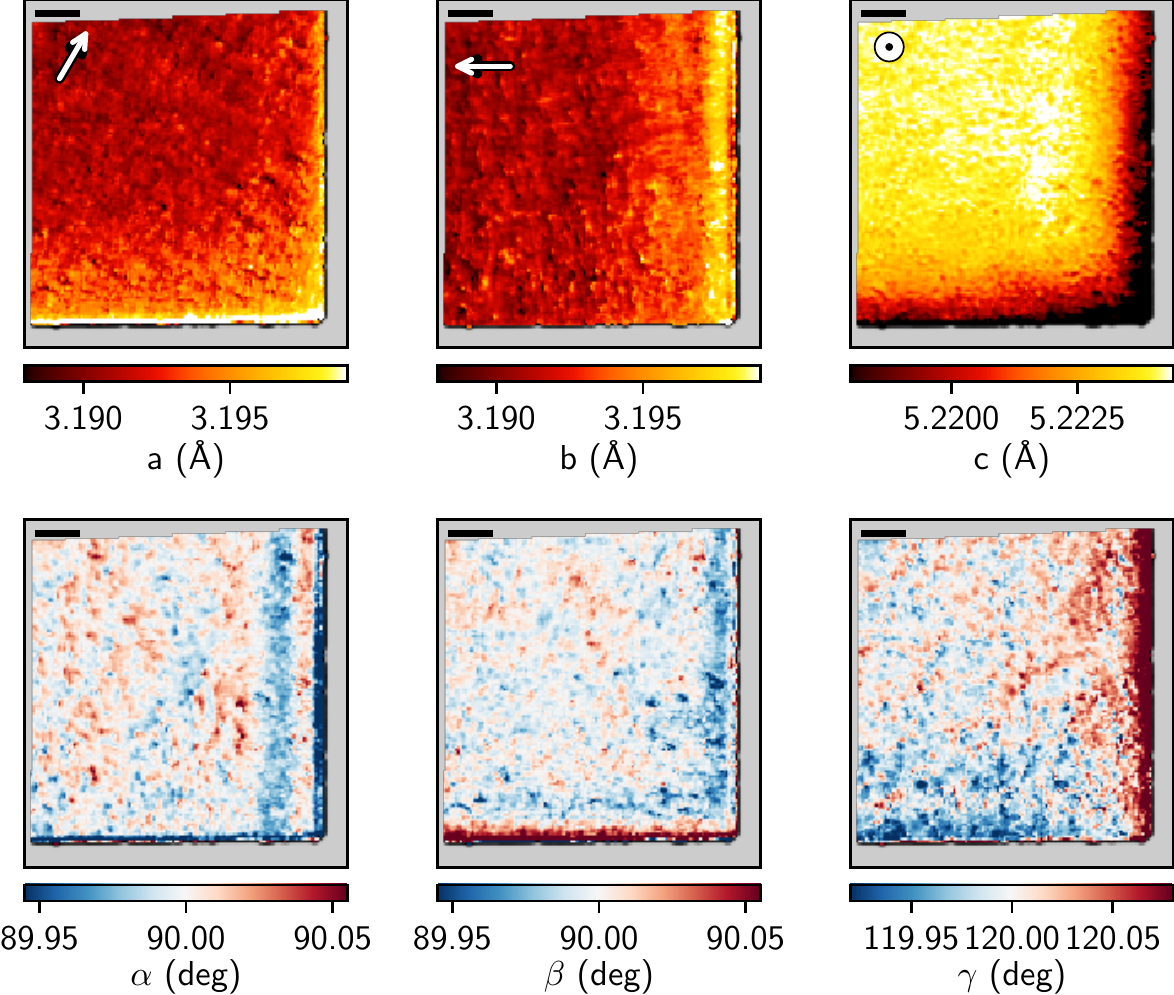}
    \caption{\label{fig:lattice_lowq} Maps of all lattice parameters for the  top InGaN layer. Arrows indicate the directions of the lattice basis vectors. Gray areas indicate missing data at the border regions due to a limited scan range either in real or reciprocal space for at least one of the three measured reflections. The scale bars correspond to $5\unit{\upmu m}$.}
\end{figure}

A more straightforward procedure can be used for the given set of three non-coplanar reflections. Based on the reciprocal lattice vectors $\mathbf Q_{0004}$, $\mathbf Q_{10\bar 13}$ and $\mathbf Q_{0\bar 1 13}$, we can calculate the basis vectors of the reciprocal lattice, $\mathbf b_1, \mathbf b_2$ and $\mathbf b_3$, via
\begin{align*}
    \label{eq:qvectors}
    \mathbf b_3  &=  \mathbf Q_{0004} / 4\\
    \mathbf b_1  &=  \mathbf Q_{10\bar 13} - 3 \mathbf b_3  \\
    \mathbf b_2  &=  -(\mathbf Q_{0\bar 113} - 3 \mathbf b_3).
\end{align*}
The real space basis vectors are then commonly obtained via $ \mathbf a_i = {2 \pi} V  \mathbf b_j \times \mathbf b_k $ with $V$ being the unit cell volume and $(i,j,k)=(1,2,3)$ and its cyclic permutations. From these vectors, the six lattice parameters can be readily calculated by all combinations of their scalar products which represent the metric tensor. They are shown for the top InGaN layer in \Fig{fig:lattice_lowq}.

Now, in order to derive rotation and strain of the lattice, we need to define a reference unit cell ($\mathbf a_1^0, \mathbf a_2^0, \mathbf a_3^0$) based, for instance, on the sample average or on that of a reference compound (here: GaN) or substrate. At any surface position, the measured unit cell (averaged over the beam footprint) is then a linear transformation $\mathbf T$ of the reference
\begin{equation}
\mathbf a_i = T_{ij} \mathbf a^0_j.
\end{equation}
Thus, we can obtain the transformation matrix $T_{ij}$ by writing the local basis vectors in matrix form and using the inverse of the reference system:
\begin{equation}
\mathbf{ T } = (\mathbf a_1, \mathbf a_2, \mathbf a_3)
\left(\mathbf a_1^0, \mathbf a_2^0, \mathbf a_3^0\right)^{-1}.
\end{equation}
From this transformation matrix, we can moreover compute the local rotation and strain tensors by means of \textit{polar decomposition} into an orthogonal rotation matrix $\mathbf U$ and a symmetric matrix $\mathbf P$ such that $\mathbf T = \mathbf U\mathbf P$. The matrix $\mathbf P$ then relates to the total strain $f_{ij}$ via 
\begin{equation}
    P_{ij} = \delta_{ij} + f_{ij}.
\end{equation}
The rotation matrix $\mathbf U$ can in general be factored into three matrices describing the rotation about any axes of the Cartesian system and thus be converted into Eulerian angles \cite{eberly2008euler}. However, in most cases, the changes in lattice orientation are very small and can be described by infinitesimal rotation that does not change the orientation of the main axes. In that case, the local lattice rotation $\omega_{ij}$ is obtained via
\begin{equation}
    U_{ij} = \delta_{ij} + \omega_{ij},
\end{equation}
where $\Delta \phi = \omega_{xy}$, $\Delta \chi = \omega_{yz}$ and $\Delta \eta = \omega_{zx}$.

\begin{figure}
    \includegraphics[width=\columnwidth]{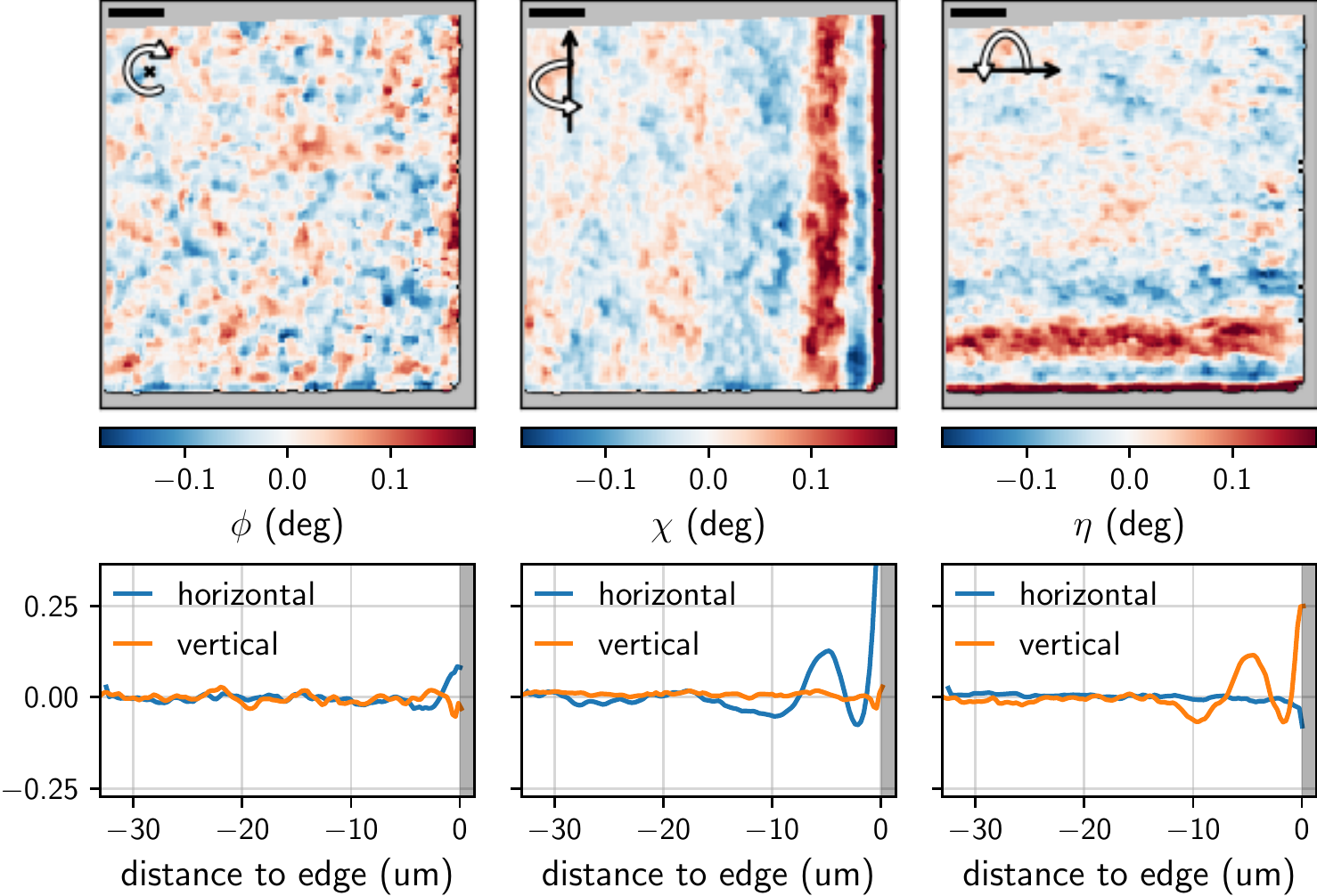}
    \caption{\label{fig:tilt_components} Rotation components $\Delta\varphi$ (yaw, $\omega_{xy}$), $\Delta\chi$ (roll, $\omega_{yz}$) and $\Delta\eta$ (pitch, $\omega_{xz}$) for the top InGaN layer. The bottom row shows projections along both sample axes. Strong tilts can be seen at the edges of the sample which is connected to a decaying waviness towards the center of the pad. Gray areas indicate missing data at the border regions due to a limited scan range either in real or reciprocal space. The scale bars correspond to $5\unit{\upmu m}$.}
\end{figure}
\Fig{fig:tilt_components} illustrates the obtained orientational variation of the crystal lattice in terms of these three angles. Here, we only show results for the regrown top InGaN layer, because maps of the two layers are almost identical and hardly any differences are visible by the eye (see Fig.~3 of the Supplementary Information). According to the RSM ranges, we can detect tilts up to approximately 1\,degree. It can be seen that the lattice is strongly tilted outwards near the edges related to the relaxation of strain. This is followed by a decaying undulation when going towards the center of the pad. The angle $\varphi$ corresponds to the twist of unit cells and does not show this undulation but also increases towards the edges of the mesa.

\subsection{Separating elastic strain and alloy composition}

The components of the total strain $f_{ij}$ consist of two contributions. The change of indium concentration in a volume element gives rise to its homogeneous expansion, provided this volume element is not restricted by the surrounding material. The respective strain $\varepsilon_{ij}^*$, called eigenstrain or intrinsic strain, does not cause stress by itself. Yet, an inhomogeneous eigenstrain causes stress, and hence elastic strain $\varepsilon_{ij}$, since the neighbor volume elements restrict each other. Other lattice defects, particularly dislocations, also contribute to the elastic strain and stress. Thus, the total strain $\mathbf f$ can be presented by the sum 
\begin{equation}
\label{eq:totalstrain}
    f_{ij} = \varepsilon^*_{ij} + \varepsilon_{ij}
\end{equation}
of the eigenstrain $\boldsymbol \varepsilon^* $ due to the indium concentration variation and the elastic strain $\boldsymbol \varepsilon $ that gives rise to stress $\boldsymbol \sigma $ via Hooke's law:
\begin{equation}
    \label{eq:strainstress}
    \sigma_{ij} = C_{ijkl} \varepsilon_{kl},
\end{equation}
where $C_{ijkl}$ is the elastic stiffness tensor. Both $\boldsymbol \sigma$ and $\boldsymbol \varepsilon$ are symmetric rank 2 tensors. Their diagonal and off-diagonal components represent normal and shear stress/strain, respectively. In the wurtzite structure of InGaN, the stiffness tensor has only 5 independent components  and its symmetry is such that there is no cross-talk between shear and normal strain \cite{Xie2012}. We assume that this symmetry is maintained for small deformations of the unit cells. 

\begin{figure}
    \includegraphics[width=1\columnwidth]{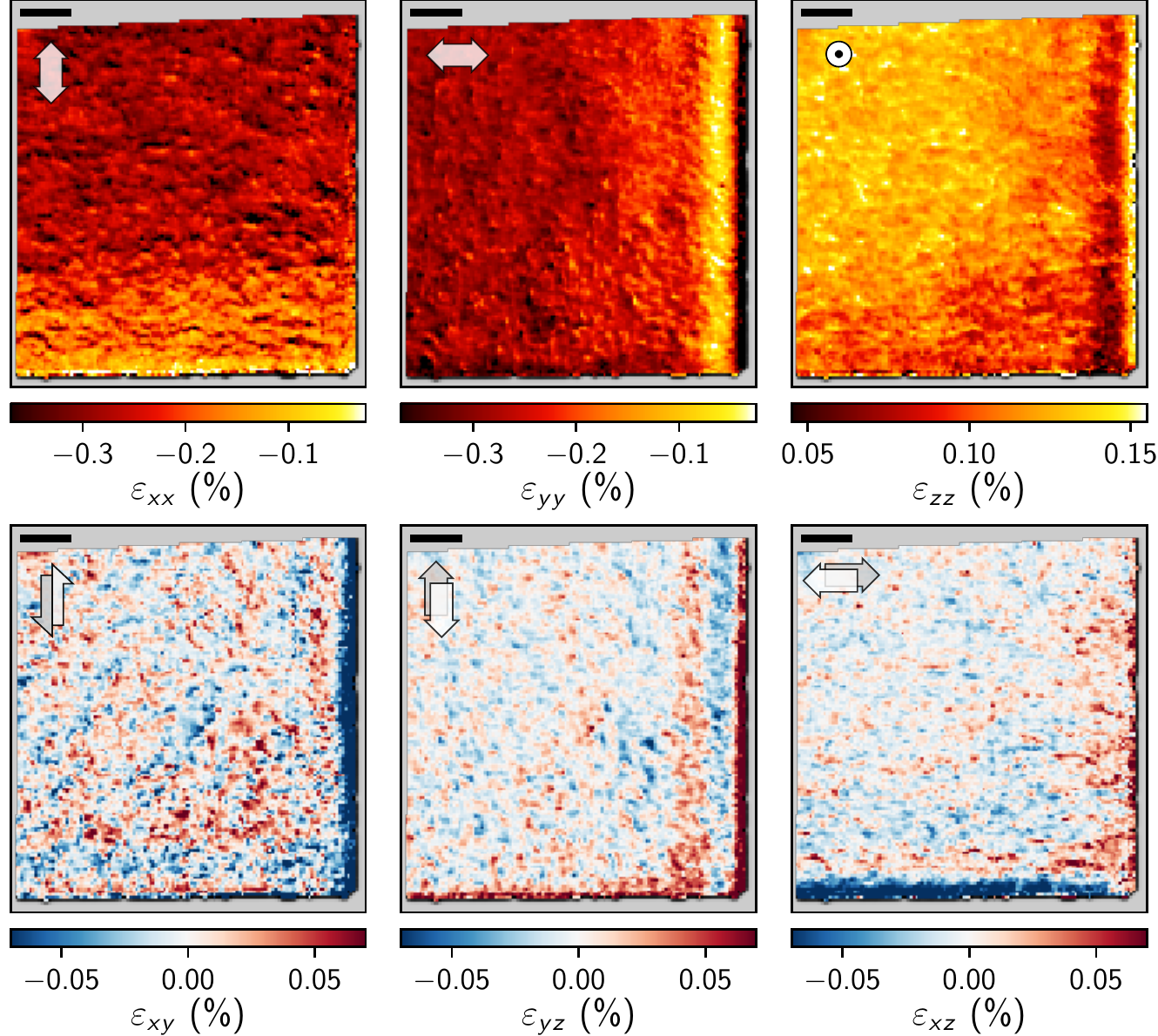}
    \caption{\label{fig:straintensor} The normal (top row) and shear (bottom row) components of elastic strain for the InGaNOS seed layer. Directions of strains are indicated by arrows. Scale bars correspond to $5\unit{\upmu m}$. }
\end{figure}

\begin{figure}
    \includegraphics[width=1\columnwidth]{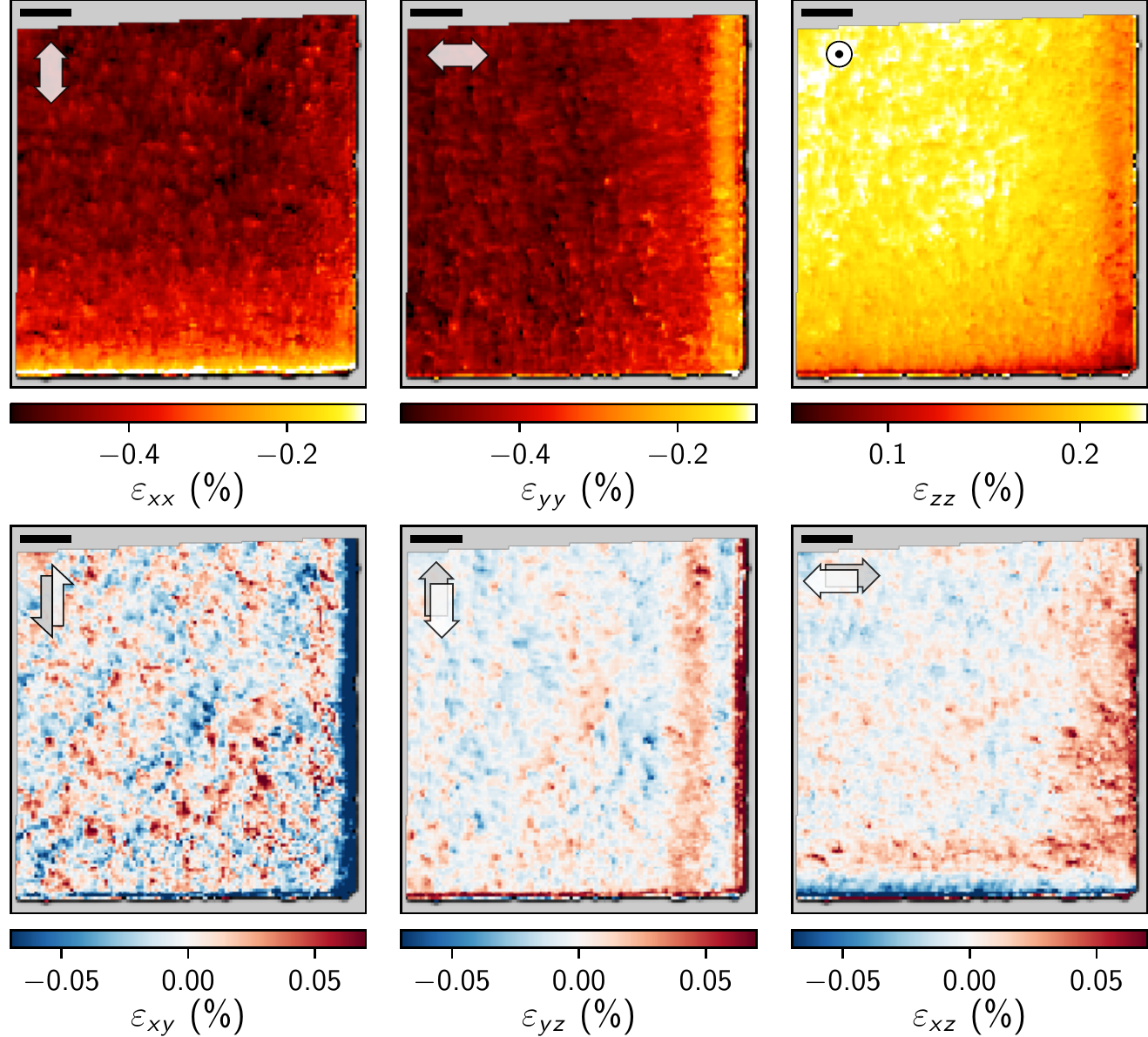}
    \caption{\label{fig:straintensor2} The normal (top row) and shear (bottom row) components of elastic strain for the regrown InGaN top layer. Directions of strains are indicated by arrows. Scale bars correspond to $5\unit{\upmu m}$.}
\end{figure}

\begin{figure}
    \includegraphics[width=0.7\columnwidth]{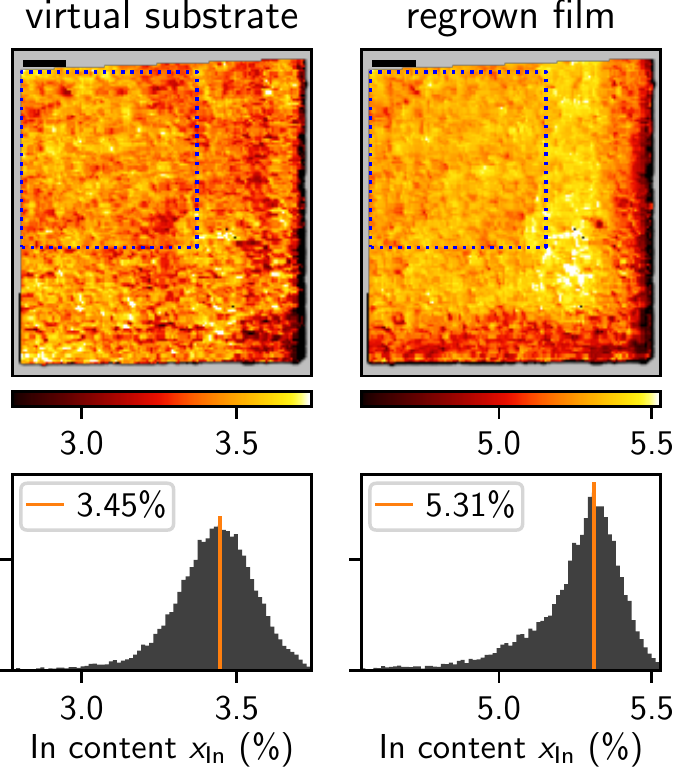}
    \caption{\label{fig:comp} Top: Maps of the indium content $x_\mathrm{In}$ in both layers: InGaNOS seed (virtual substrate, left) and the regrown InGaN layer (right). The corresponding histograms are shown in the bottom row. Scale bars correspond to $5\unit{\upmu m}$. The $20\times 20\unit{\upmu m^2}$ area far away from edges which is used for statistical analysis below is marked by a dotted blue line. }
\end{figure}

The surface of the film is free from applied forces and hence $\sigma_{zz}=0 $ at the surface. For a thin film, assuming that indium concentration does not vary over the thickness, one can assume that the condition $\sigma_{zz}=0 $ is satisfied also in the bulk of the film (plane-stress approximation).  This condition allows the determination of the indium concentration and its variation over the film.  For (0001)-grown epitaxial layers with wurtzite structure, \Eq{eq:strainstress} then yields \cite{Xie2012}
\begin{equation}
    \label{eq:biaxial}
    \sigma_{zz} = 0 = C_{13} \varepsilon_{xx} +  C_{13} \varepsilon_{yy} + C_{33} \varepsilon_{zz},
\end{equation}
where we used the Voigt notation and the symmetry of the stiffness tensor $C_{13}=C_{23}$.
We note that, on microscopic level, the in-plane isotropy $\varepsilon_{xx} = \varepsilon_{yy}$ cannot be assumed. After rearranging \Eq{eq:biaxial}, $C_{13}/C_{33}$ remains as the only unknown parameter, which can be substituted by $\nu/(1-\nu)$ with $\nu$ being the material-dependent Poisson ratio. For a given indium content $x_\mathrm{In}$ in \ce{In_xGa_{1-x}N}, $\nu(x_\mathrm{In})$ and the relaxed lattice parameters are derived by linear interpolation of values known for the binary compounds \ce{InN} and \ce{GaN}. The reference lattice parameters $\mathbf a_i^0$ and hence the eigenstrain $\varepsilon^*_{ij}$ are defined with respect to GaN:
\begin{align}
    \label{Vegard1}
    \nu (x_\mathrm{In}) &= x_\mathrm{In} \nu_\mathrm{InN} + (1-x_\mathrm{In}) \nu_\mathrm{GaN}\\
    \label{Vegard2}
    \varepsilon^*_{xx} (x_\mathrm{In}) = \varepsilon^*_{yy}(x_\mathrm{In}) &=  x_\mathrm{In}  \left( \frac {a_\mathrm{InN}} {a_\mathrm{GaN}} - 1 \right) \\
    \label{Vegard3}
    \varepsilon^*_{zz}(x_\mathrm{In}) & =  x_\mathrm{In}  \left( \frac {c_\mathrm{InN}} {c_\mathrm{GaN}} - 1 \right) \\
    \label{Vegard4} 
    \varepsilon^*_{ij} & =  0, \,  i \neq j . \nonumber
\end{align}
We use the following values from Ref.~\cite{Moram2009}: $a_\mathrm{GaN} = 3.1878\angstrom$, $c_\mathrm{GaN} = 5.185\angstrom$, $a_\mathrm{InN} = 3.538\angstrom$, $c_\mathrm{InN} = 5.703\angstrom$, $\nu_\mathrm{GaN} = 0.183$, $\nu_\mathrm{InN} = 0.272$.
%
%
%
%
Given the expression for total strain $f_{ij}$ in \Eq{eq:totalstrain} and Eqs.~(\ref{Vegard1}),  (\ref{Vegard2}), (\ref{Vegard3}), the plane-stress condition \Eq{eq:biaxial} results in an equation that is quadratic in $x_\mathrm{In}$ as the only unknown. It can be solved analytically and and has only one root in the range $0<x_\mathrm{In}<1$. The solution for every surface coordinate ($x, y$) provides spatial maps of strain and alloy composition of the layer. The former are shown in \Fig{fig:straintensor} and \Fig{fig:straintensor2} for the top InGaN layer and the InGaNOS seed layer, respectively. The composition maps for both layers are shown in \Fig{fig:comp}.

The maps in Figs.\ \ref{fig:tilt_components}--\ref{fig:comp} reveal strong changes of the lattice orientation, strain, and In content at the mesa edges (the right and the bottom edges of the maps). The in-plane strain components $\varepsilon_{xx}$ and $\varepsilon_{yy}$ in Figs.\ \ref{fig:straintensor} and \ref{fig:straintensor2} relax each at the edge where the respective strain component is normal to the edge  (i.e., $\varepsilon_{xx}$ relaxes at the bottom edge while $\varepsilon_{yy}$ relaxes at the right edge), which is expected due to an absence of geometrical restrictions. For the InGaNOS pseudo substrate (\Fig{fig:straintensor}), nearly full strain relaxation is observed for the corresponding components. As a consequence of relaxation, the out-of-plane strain $\varepsilon_{zz}$ reduces at both edges.
As mentioned before, this is linked to lattice undulations setting at a $10\unit{\upmu m}$ away from the edge (see \Fig{fig:tilt_components}). The spacial frequency of the undulations increases as the edge of the mesa is approached. Note that the maps of lattice rotation are nearly identical for both InGaN layers (see Fig.~3 of the Supplementary Information). It is not yet clear how strain relaxation leads to this buckling effect.

In \Fig{fig:comp}, one can see a reduced In-uptake in the regrown InGaN layer close to the edges of the mesa structure, which stands in contrast to what is expected based on the relief of compressive strain in these regions. A comparison to \Fig{fig:intensity}(a) shows that this loss of In is connected to an increased V-pit density. Such a reduced In concentration near V-pits has already been observed before \cite{Zoellner2019}. It may be the result of a disturbed growth such as changes in the relative diffusion of In and Ga due to the high defect density. As expected, the In distribution in the InGaNOS seed layer is hardly affected, since it was formed before the patterning took place. 

\subsection{Statistical comparison of lattice parameters}
\label{ch:stats}
\begin{figure}
    \includegraphics[width=1\columnwidth]{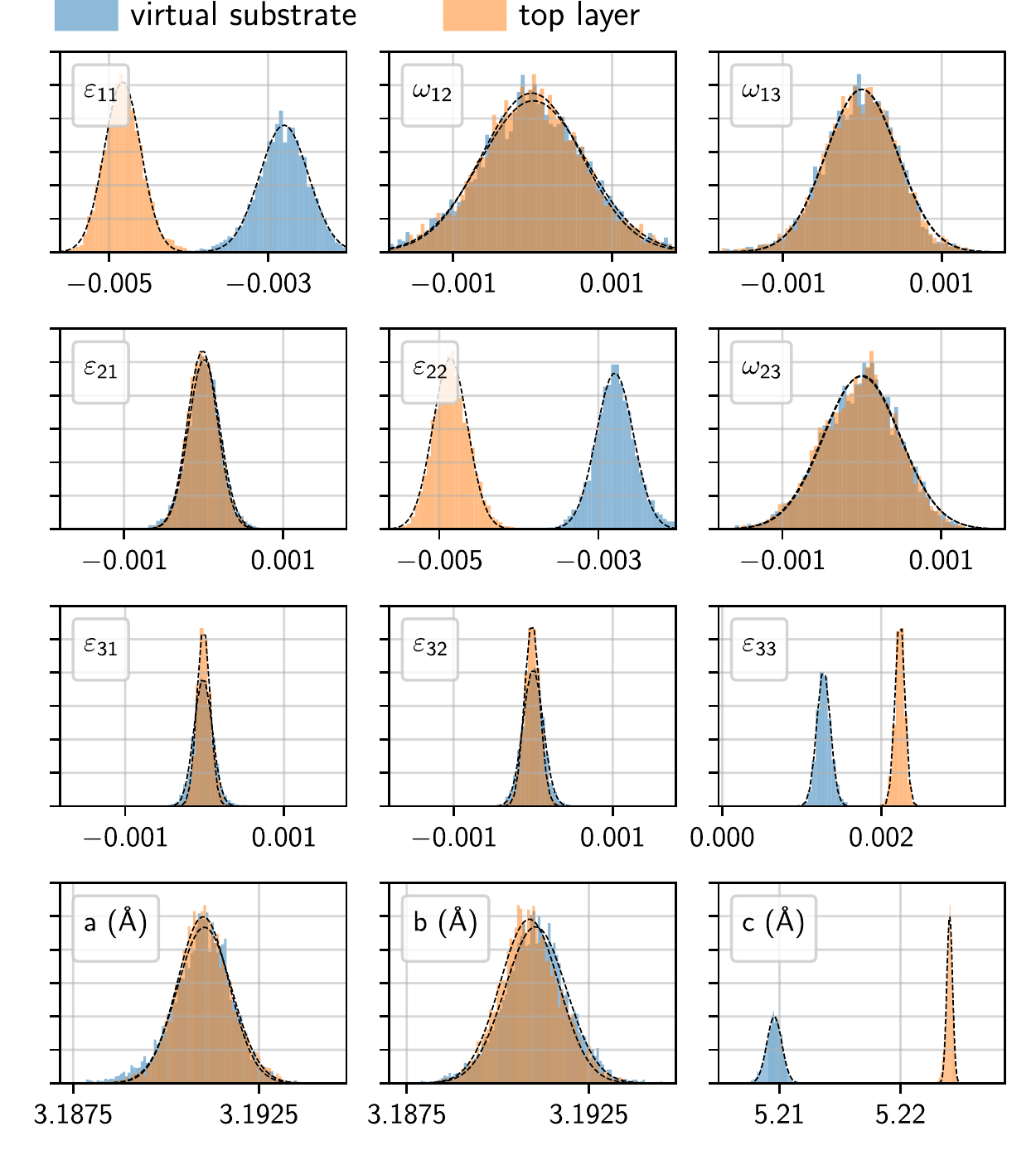}
    \caption{\label{fig:strain-histo} Histograms of elastic strain and lattice rotation in both InGaN layers derived from the maps in Figs.~\ref{fig:tilt_components}, \ref{fig:straintensor}, \ref{fig:straintensor2}. For the histograms of strain and rotation, the same ranges have been used for the abscissae to facilitate the comparison. Gaussian fits are shown by dashed lines and the mean values and standard deviations are presented in \Tab{tab:statistics}. 
    }
\end{figure}

In the following sections, we focus our analysis on a $20\times 20\unit{\upmu m^2}$ region in the bulk part of the mesa, where the edge effects are not essential (marked as dotted squares in the top left corners of the maps \Fig{fig:comp}). We first consider the statistical characteristics of the maps and then discuss possible origins of the variations in terms of the microstructure.

\Fig{fig:strain-histo} shows the histograms obtained from the maps of strain, rotation and lattice parameters for both bottom and top layer. The variations of lattice rotations $\omega_{ij}$ are larger, and shear strain is smaller compared to the normal strain components. The narrower In distribution of the top InGaN film (see \Fig{fig:comp}) leads to a narrower distribution of the out-of-plane lattice parameter $c$. The distributions of the in-plane lattice parameters $a$ and $b$ of the two layers coincide, since the layers are epitaxially linked. The distributions of the normal strains in the plane therefore reflect the variations of the indium content. The histograms can be well described by normal distributions as shown in Fig.\ \ref{fig:strain-histo} by dotted lines. Their mean values and standard deviations are presented in \Tab{tab:statistics}.
\begin{table*}
\caption{\label{tab:statistics} Statistical parameters of the virtual substrate (InGaNOS) and the regrown InGaN layer. Only a $20\times 20 \unit{\upmu m}^2$ sample area omitting the edges of the mesa (top left corner) is taken into account (marked by dotted squares in \Fig{fig:comp}). The autocorrelation lengths are defined as the distance where the radial autocorrelation function drops to half of its value at zero distance (see Figs.~1 and 2 of the Supplementary Information). The Pearson $r$ value characterizes the cross-correlation between InGaNOS seed and top InGaN layer.}
    \begin{ruledtabular}\small
        \begin{tabular}[c]{crrrrr}
            Quantity & \multicolumn{2}{c}{mean value $\pm$ standard deviation} & \multicolumn{2}{c}{autocorrelation length (nm)} & cross-correlation \\
            (unit)& \multicolumn{1}{c}{InGaNOS seed} &  \multicolumn{1}{c}{top layer} &  \multicolumn{1}{c}{InGaNOS seed} &  \multicolumn{1}{c}{top layer}  & Pearson $r$\\
            \hline
            $\alpha$ (deg)       & $90 \pm 0.014$        & $90 \pm 0.011$        & 722 & 895 & 0.61\\
            $\beta$ (deg)        & $90 \pm 0.013$        & $90 \pm 0.009$        & 635 & 750 & 0.56\\
            $\gamma$ (deg)       & $120 \pm 0.020$       & $120 \pm 0.019$       & 651 & 711 & 0.55\\
            $a\ (\angstrom)$  & $3.1910 \pm 0.0007$   & $3.1910 \pm 0.0007$   & 587 & 667 & 0.56\\
            $b\ (\angstrom)$  & $3.1909 \pm 0.0009$   & $3.1908 \pm 0.0008$   & 741 & 859 & 0.64\\
            $c\ (\angstrom)$  & $5.2096 \pm 0.0006$   & $5.2241 \pm 0.0002$   & 974 & 609 & 0.09\\
            $\omega_{xy}=\Delta\varphi$ (mrad)     & $0.01 \pm 0.60$      & $-0.01 \pm 0.63$     & 1104 & 1169 & 0.96\\
            $\omega_{yz}=\Delta\chi$ (mrad)        & $0.00 \pm 0.48$       & $-0.01 \pm 0.48$     & 1460 & 1428 & 0.96\\
            $\omega_{zx}=\Delta\eta$ (mrad)        & $0.00 \pm 0.45$       & $-0.01 \pm 0.45$     & 1395 & 1512 & 0.92\\
            $x$ (\%)             & $3.459 \pm 0.098$     & $5.305 \pm 0.063$     & 805  & 862 & 0.56\\
            $\varepsilon_{xx}\, (10^{-3})$   & $-2.80 \pm 0.31$    & $-4.83 \pm 0.22$    & 711  & 758 & 0.39\\
            $\varepsilon_{yy}\, (10^{-3})$   & $-2.79 \pm 0.23$    & $-4.86 \pm 0.22$    & 744  & 837 & 0.61\\
            $\varepsilon_{zz}\, (10^{-3})$   & $1.28 \pm 0.08$     & $2.24  \pm 0.06$   & 766  & 751 & 0.33\\
            $\varepsilon_{xy}\, (10^{-3})$   & $0.01 \pm 0.19$     & $0.02  \pm 0.19$   & 648  & 731 & 0.62\\
            $\varepsilon_{xz}\, (10^{-3})$   & $0.02 \pm 0.11$     & $0.01  \pm 0.08$   & 646  & 798 & 0.62\\
            $\varepsilon_{yz}\, (10^{-3})$   & $0.01 \pm 0.13$     & $0.01  \pm 0.09$   & 722  & 895 & 0.61\\
        \end{tabular}
    \end{ruledtabular}
\end{table*}

One can see by eye that the maps of strain components (Figs.~\ref{fig:straintensor} and \ref{fig:straintensor2}) exhibit sharper features compared to the maps of rotation (\Fig{fig:tilt_components}), which is not reflected in the histograms. To quantify this effect, it is useful to look at autocorrelation functions of the respective maps. The autocorrelations for the top InGaN layer are shown in Fig.~\ref{fig:ProbAutocorr}. The other autocorrelation functions for both layers are presented in Figs.~1 and 2 of the Supplementary Information. We define a characteristic length scale as a distance where the radial correlation function (after azimuthal integration) drops to half of its maximum value. These lengths are given in \Tab{tab:statistics} next to the other statistical parameters. One should keep in mind, that the experimental resolution of $\approx 200\dots 300\unit{nm}$ poses a lower boundary to the autocorrelation lengths.

In general, we can see variations on the micrometer scale in all of the experimental maps which has been observed in \IGN{} films before \cite{Zoellner2019, Butt2018}. It also appears that the lattice parameters, strains and rotations of the top InGaN layer vary on longer length scales compared to the seed layer. That might be a result of the averaging over a twice larger layer thickness, with a certain lateral averaging due to an inclination of the \xray{} beam with respect to the layer plane. The only clear exception is the out-of-plane lattice parameter $c$, which instead varies on smaller distances in the top layer due to a narrower distribution of In in the considered part of the mesa. The data in \Tab{tab:statistics} also reveals the significantly (approximately twice) larger autocorrelation lengths for the rotation components in comparison to the strain components. The lattice rotation is usually related to threading dislocations of edge or screw type in epitaxial GaN layers \cite{Moram2009}. The high degree of correlation of the maps of rotation components in top and bottom layers thus show that the dislocations are mostly inherited from the virtual substrate. 


While the distributions of rotations with the in-plane rotation axes (tilt, $\omega_{xz}$ and $\omega_{yz}$) have nearly identical values of both width and correlation length, they significantly differ from rotations about the surface normal (twist, $\omega_{xy}$). This fits well into the picture of threading dislocations, since threading edge ($a$-type) dislocations contribute to the twist while threading screw ($c$-type) dislocations contribute to the tilt components \cite{anderson2017theory,kosevich1979crystal}.
A quantitative analysis of the strain distributions and the autocorrelation functions is presented in Sec.\  \ref{subsec:MCanalysis} below. In a simplified picture according to Ref.~\cite{Metzger1998} for randomly distributed dislocations, taking the standard deviations of the rotation parameters in \Tab{tab:statistics}, we obtain densities of  $1.2\times 10^8\,\mathrm{cm}^{-2}$ and  $5.1\times 10^8\,\mathrm{cm}^{-2}$ for screw and edge dislocations, respectively. Since we have assumed normal distributions (see \Fig{fig:strain-histo}), we introduce a factor of $\sqrt{2\pi}$ to convert the standard deviation into the integral breadth. Comparing with \Fig{fig:intensity}(b), we can see that the V-pit density is on the same order of magnitude in the considered upper left corner of the sample. Calculating histograms of smaller sample regions, maps of local dislocation densities can be obtained. 



\subsection{Elastic strain due to inhomogeneous indium composition}

An inhomogeneous distribution of In in the film gives rise to elastic strain. The aim of this section is to evaluate this strain and subtract it from the strain measured by \xray{} diffraction. The difference is attributed to threading dislocations and considered in the next section.

A homogeneous In distribution in GaN results in a homogeneous lattice expansion, similarly to thermal expansion. Moreover, since the relative changes of \emph{a} and \emph{c} lattice parameters between GaN and InN are almost identical ($\Delta a/a\approx\Delta c/c\approx10\%$, here $\Delta a$ and $\Delta c$ are the differences between the respective lattice parameters of InN and GaN), the strain due to a homogeneous indium concentration is $\varepsilon_{ij}^{*}=\varepsilon^{*}\delta_{ij}$ (where $\delta_{ij}$ is the Kronecker delta,  $\varepsilon^{*}=x_{\mathrm{In}}\Delta a/a$ and $x_{\mathrm{In}}$ is the indium concentration, see \Fig{fig:comp}). This strain (eigenstrain or intrinsic strain, in terminology of the theory of internal stresses) is equivalent to a thermal strain $\varepsilon^{*}=\alpha \Delta T$, where $\alpha$ is thermal expansion coefficient and $\Delta T$ temperature difference. The strain $\varepsilon^{*}$ describes a homogeneous crystal expansion which itself does not cause elastic strain and stress, provided that a piece of material with the strain $\varepsilon^{*}$ is not constrained by the surrounding material.

The constraints imposed by the continuity of the material for an inhomogeneous indium distribution, i.e., for $\varepsilon^{*}(\mathbf{r})$ varying in space, give rise to an additional elastic strain, which can be found by solving the elastic equilibrium equations. The solution of the respective thermoelastic problem for the strain due to an inhomogeneous temperature distribution in a thin plate is well known \cite{melan_parkus53}. Hexagonal symmetry of GaN in (0001) plane gives rise to the transverse elastic isotropy, so that the isotropic solution can be used with the Poisson ratio $\nu=C_{12}/(C_{11}+C_{12})$ (using Voigt notation). For the elastic moduli of GaN \cite{polian96}, we get $\nu\approx0.27$. Hence, it remains to reformulate the thermoelastic solution in our notation.

The solution \cite{melan_parkus53} is expressed through the thermoelastic potential $\Psi$ satisfying the equation 
\begin{equation}
\frac{\partial^{2}\Psi}{\partial x^{2}}+\frac{\partial^{2}\Psi}{\partial y^{2}}=(1+\nu)\varepsilon^{*}(x,y).\label{eq:In-1}
\end{equation}
The displacements are $u_{x}=\partial\Psi/\partial x$, $u_{y}=\partial\Psi/\partial y$, and the components of the elastic strain due to an inhomogeneous In distribution $\varepsilon_{ij}^{\mathrm{In}}=\frac{1}{2}(\partial u_{i}/\partial x_{j}+\partial u_{j}/\partial x_{i})$ are 
\begin{equation}
    \varepsilon_{xx}^{\mathrm{In}}=\frac{\partial^{2}\Psi}{\partial x^{2}},\,\,\,\varepsilon_{yy}^{\mathrm{In}}=\frac{\partial^{2}\Psi}{\partial y^{2}},\,\,\,\varepsilon_{xy}^{\mathrm{In}}=\frac{\partial^{2}\Psi}{\partial x\partial y},\,\,\,\varepsilon_{zz}^{\mathrm{In}}=(1+\nu)\varepsilon^{*}.\label{eq:In-3}
\end{equation}

The average of the eigenstrain $\varepsilon^{*}(x,y)$ over the plate produces a homogeneous expansion and can be subtracted. Hence, we consider in Eq.~(\ref{eq:In-1}) the eigenstrain with zero average value. The effect of the borders of the plate is restricted, according to the Saint-Venant's principle, to a stripe of the width comparable with the characteristic length of the fluctuations in the In concentration. 
Since our area of interest is well away from the borders, we neglect the boundary conditions to Eq.~(\ref{eq:In-1}) and solve it by Fourier transformation of the eigenstrain 
\begin{equation}
    \varepsilon^{*}(\mathbf{r})=\int\varepsilon_{\mathbf{k}}^{*}e^{-2\pi i\mathbf{k}\cdot\mathbf{r}}d\mathbf{k},\label{eq:In-4}
\end{equation}
where $\mathbf{r}$ and $\mathbf{k}$ are the two-dimensional radius
vector and wave vector, respectively. Solving Eq.~(\ref{eq:In-1})
and substituting the solution in Eq.~(\ref{eq:In-3}), we get for
the in-plane components of the strain due to an inhomogeneous
In distribution in a thin plate
\begin{equation}
\varepsilon_{ij}^{\mathrm{In}}(\mathbf{r})=(1+\nu)\int\frac{k_{i}k_{j}}{k^{2}}\varepsilon_{\mathbf{k}}^{*}e^{-2\pi i\mathbf{k}\cdot\mathbf{r}}d\mathbf{k}\,\,\,\,\,(i,j=1,2).\label{eq:In-5}
\end{equation}
The $\varepsilon_{zz}^{\mathrm{In}}$ component is calculated directly
by the last expression Eq.~(\ref{eq:In-3}).

\begin{figure*}
    \includegraphics[width=1\textwidth]{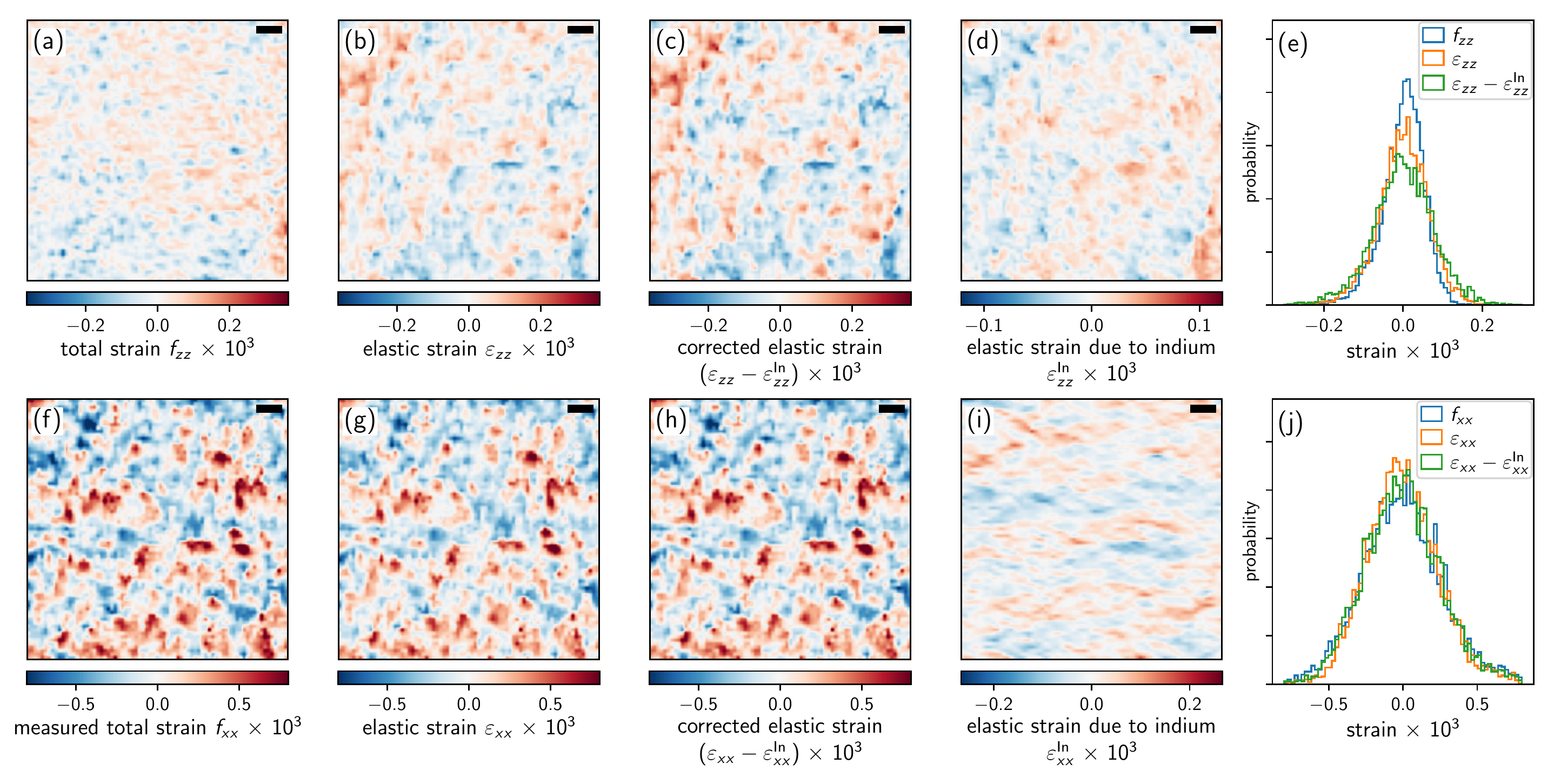}
    \caption{(a) Total strain $f_{zz}$ as measured by \xray{} diffraction, (b) elastic strain  $\varepsilon_{zz}$ after subtraction of the eigenstrain $\varepsilon^{*}$, (c) corrected elastic strain $\varepsilon_{zz}-\varepsilon_{zz}^{\mathrm{In}}$ after subtracting the elastic strain $\varepsilon_{zz}^{\mathrm{In}}$ (d) that is due to inhomogeneous In concentration, (e) histograms of the $zz$ components of total, elastic and corrected strain. (f-j) the same data as (a-e) for the $xx$ component of the strain tensors. The size of all maps is $20\times20\unit{\upmu m^{2}}$.}
\label{fig:In-maps} 
\end{figure*}

We show below that the $f_{zz}$ component of total strain as measured by \xray{} diffraction contains a substantial contribution due to the inhomogeneous In distribution, while the effect of In on the in-plane strain components is minor. Hence, we begin the analysis with the $\varepsilon_{zz}^{\mathrm{In}}$ component. \Fig{fig:In-maps}(a) is the map $f_{zz}$ measured by \xray{} diffraction (the average over the map is subtracted). \Fig{fig:In-maps}(b) shows the elastic strain $\varepsilon_{zz}$ which is the difference between the total strain $f_{zz}$ and the eigenstrain $\varepsilon^{*}$ obtained from the In concentration map $x_\mathrm{In}$ (see \Fig{fig:comp}). We subtract the elastic strain due to inhomogeneous indium concentration $\varepsilon_{zz}^{\mathrm{In}}$ which is calculated by Eq.~(\ref{eq:In-3}) and shown in \Fig{fig:In-maps}(d). The remaining elastic strain is shown for comparison in \Fig{fig:In-maps}(c). \Fig{fig:In-maps}(e) compares the probability distributions of the measured total strain $f_{zz}$, the derived elastic strain $\varepsilon_{zz}$ and the remaining elastic strain after subtracting the elastic strain $\varepsilon_{zz}^{\mathrm{In}}$ due to an inhomogeneous In distribution. This latter quantity is attributed to threading dislocations and will be discussed below.

Figs.~\ref{fig:In-maps}(f-i) show the same data for the in-plane components of strain $f_{xx}$ and $\varepsilon_{xx}$. One can see that, in this case, the correction is negligible. This is confirmed by the histograms in Fig.~\ref{fig:In-maps}(j) of the probability distributions of the prior and after the correction to the strain due to In. The effect on the other in-plain strain components $\varepsilon_{xy}$ and $\varepsilon_{yy}$ is also negligible compared with the accuracy of the measurements. Hence, the measured maps of the in-plane strain components are due to the strain from sources different from the In distribution.

\subsection{Elastic strain due to threading dislocations}
\label{subsec:MCanalysis}
We attribute the remaining strain in the film to threading dislocations crossing the film along its normal. For the present density of dislocations, the resolution of the measurements is not sufficient to resolve individual threading dislocations. However, it is possible to analyze the linear superpositions of their strain and rotation fields which are a result of the measurement. 
In the Supporting Information, we derive explicit formulas for all components of the strain and rotation tensors due to dislocations crossing the film along its normal, taking account of the elastic stress relaxation on the free surfaces of the film. These fields consist of two contributions: the long range ($\propto\rho^{-1}$) field provides the main contribution to the probability distributions and the correlation functions described below, while further relaxation terms decay faster with the distance $\rho$ from the dislocation line and give rise to only little correction of the results. The long range parts of the strain components are directly related to the dislocation strains and rotations in an infinite medium. 

An edge dislocation in an infinite medium gives rise to strain in the plane perpendicular to the dislocation line. When a thin film is cut perpendicular to the dislocation line, the components of the strain at distances $\rho$ exceeding the film thickness are \begin{eqnarray}
    \varepsilon_{\rho\rho} & = & \varepsilon_{\phi\phi}=-\frac{b_{x}}{4\pi}(1-\nu)\frac{\sin\phi}{\rho},\label{eq:e1}\\
    \varepsilon_{\rho\phi} & = & \frac{b_{x}}{4\pi}(1+\nu)\frac{\cos\phi}{\rho},\,\,\,\varepsilon_{zz}=\nu\frac{b_{x}}{2\pi}\frac{\sin\phi}{\rho},\nonumber 
\end{eqnarray}
where $b_{x}$ is the Burgers vector. Here we describe the components in cylindrical coordinates (see Supplementary Information). The in-plane strain components in the film differ from the respective expressions for the dislocation strain in the infinite medium by a substitution of the Poisson ratio $\nu$ with $\nu/(1+\nu)$ (the plane stress solution). The strain $\varepsilon_{zz}$ arises to provide zero normal stress, $\sigma_{zz}\propto(1-\nu)\varepsilon_{zz}+\nu(\varepsilon_{\rho\rho}+\varepsilon_{\phi\phi})=0$.
The strains $\varepsilon_{\rho z}$ and $\varepsilon_{\phi z}$ are zero in the infinite medium and decay faster than $\rho^{-1}$ in the film. The in-plane rotation due to an edge dislocation is the same as in the infinite medium, 
\begin{equation}
    \omega_{\rho\phi}=\frac{b_{x}}{2\pi}\frac{\cos\phi}{\rho},\label{eq:e2}
\end{equation}
while $\omega_{\rho z}$ and $\omega_{\phi z}$ decay faster than $\rho^{-1}$.

The displacement field of a screw dislocation in an infinite medium $u_{z}=(b_{z}/2\pi)\phi$, where $b_{z}$ is the Burgers vector, gives rise to the strain and rotation $\varepsilon_{\phi z}=\omega_{\phi z}=b_{z}/4\pi\rho$. In a thin film, the long-range strain $\varepsilon_{\phi z}$ is zero to provide stress-free boundary condition. As a result, the rotation in the film 
\begin{equation}
    \omega_{\phi z}=\frac{b_{z}}{2\pi\rho}\label{eq:e3}
\end{equation}
is two times larger than it is in the infinite medium, and remains the only long-range component of the strain and rotation tensors; all other components decay faster than $\rho^{-1}$. 
The corrections to the strain and rotation tensors at the distances from the dislocation line comparable with the film thickness are derived in the Supporting Information and provide only small corrections to the strain and rotation fields.


Threading dislocations in GaN are correlated, to reduce elastic energy due to their slowly decaying strain fields \cite{kaganer05GaN}. These correlations can be modeled by pairs of dislocations with opposite Burgers vectors. When  the mean distance $R$ between dislocations in the pairs exceeds the distance between dislocations in the crystal $r_{d}=1/\sqrt{\varrho}$, where $\varrho$ is the density of threading dislocations (i.e., the density of the pairs is $\varrho/2$), the pairs overlap \cite{kaganer10acta}. The range of the correlations can be described by the dimensionless parameter $M=R/r_{d}=R\sqrt{\varrho},$ introduced by Wilkens \cite{wilkens70pss,wilkens70nbs,wilkens76}. In his model of the `restrictedly random dislocation distribution', the crystal is divided in cells, each cell containing $M$ dislocations with the total Burgers vector equal to zero. Modeling by dislocation pairs, with the parameter $M$ defining the mean distance between dislocations in the pair, is more convenient and gives very close diffraction profiles \cite{kaganer10acta}. Further details of the Monte Carlo modeling of the dislocation arrays are given in the Appendix \ref{appendix}.


\begin{figure}
    \includegraphics[width=\columnwidth]{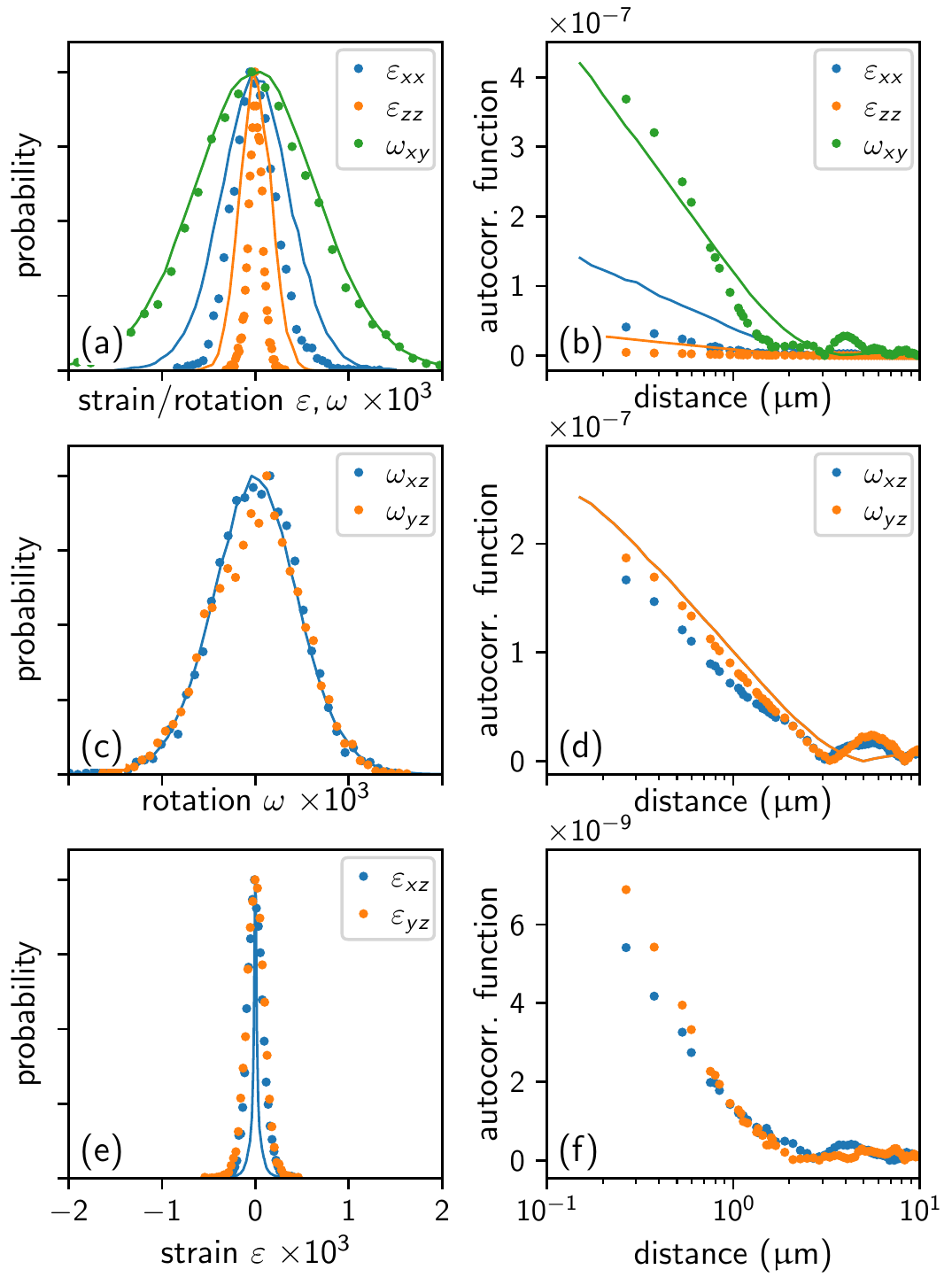}
    \caption{Comparison of histograms and autocorrelation functions of the measured and simulated distributions of strain $\varepsilon_{ij}$ and rotation $\omega_{ij}$. (a,b) The in-plane strain and rotation components, as well as the normal strain $\varepsilon_{zz}$, that are attributed to edge threading dislocations with the density of $2\times 10^9\,\mathrm{cm}^{-2}$. (c,d) the tilt components $\omega_{xz},\omega_{yz}$ attributed to screw threding dislocations with the density of $4\times 10^8\,\mathrm{cm}^{-2}$. The dislocations are modeled by random dislocation pairs with the Wilkens' parameters $M=10$ for edge amd $M=7$ for screw dislocations to optimize the correlation lengths. The distances between dislocations in the pairs are described by a lognormal distribution with a standard deviation set to half of the mean distance between dislocations in the pairs. (e,f) The shear strain components $\varepsilon_{xz},\varepsilon_{yz}$ that are expected to be zero in the plane stress approximation. Note that the scale in (f) is different from that in (b,d). The measured data are shown by circles, whereas Monte Carlo simulations for edge and screw dislocations are shown by lines.}
\label{fig:ProbAutocorr}
\end{figure}

We adjusted the parameters $\varrho$ and $M$ for both threading screw and threading edge dislocations so that the histograms and autocorrelations obtained from Monte Carlo modeling of the  the strain and rotation fields agree with the experiment. We primarily fit the rotation components, since they are larger than strains. The in-plane rotation $\omega_{xy}$, as well as the in-plane strain components $\varepsilon_{xx},\varepsilon_{xy},\varepsilon_{yy}$ and the strain normal to the plane $\varepsilon_{zz}$, are attributed to edge dislocations and presented in Fig.\ \ref{fig:ProbAutocorr}(a,b). The tilt components $\omega_{xz}$ and $\omega_{yz}$ are attributed to screw dislocations and presented in Fig.\ \ref{fig:ProbAutocorr}(c,d). The obtained densities of screw and edge dislocations are $4\times 10^8\,\mathrm{cm}^{-2}$ and  $2\times 10^9\,\mathrm{cm}^{-2}$ , respectively, 
about four times larger than those obtained above (see Sec.\ \ref{ch:stats}) following Ref.~\cite{Metzger1998}. The correlation parameters are found to be $M=10$ for edge and $M=7$ for screw dislocations. Hence, the respective distances of the screening of the strain fields of dislocations by surrounding dislocations $R=M/\sqrt{\varrho}$ are 2.2~µm and 3.5~µm for edge and screw dislocations. These screening distances are directly seen in Figs.\ \ref{fig:ProbAutocorr}(b) and \ref{fig:ProbAutocorr}(d) as the intersections of the straight lines (in the lin-log scale) with the ordinate axis.

The shear strain components $\varepsilon_{xz}$ and $\varepsilon_{yz}$ (\Fig{fig:ProbAutocorr}(e,f)) are 
zero in the plane-stress approximation. The Monte Carlo calculation in Fig.\ \ref{fig:ProbAutocorr}(e) is performed using three-dimensional strain fields of the dislocations in the film derived in Supporting Information. Both edge and screw dislocations with the densities and correlations obtained above are included in the calculation. One can see from Fig.\ \ref{fig:histograms}(a,c) in Appendix \ref{appendix}, that at equal densities, screw dislocations provide larger contribution to $\varepsilon_{xz}$ and $\varepsilon_{xz}$ compared to edge dislocations. As a result, screw dislocations with smaller density and edge dislocations with larger density give comparable contributions to the curve. 

The observed variations of $\varepsilon_{xz}$ and $\varepsilon_{yz}$  in the experimental histograms in \Fig{fig:ProbAutocorr}(e) are larger than the simulated ones, but remain significantly smaller than these of the other strain or rotation components. They can be attributed either to experimental error or to limitations of the model. Particularly, we did not consider potential variations of the indium content over the depth and the residual stress on the film due to the bonding on the handling wafer.


\section{Conclusions}
For the first time, we demonstrate that scanning \xray{} diffraction microscopy (SXDM) provides quantitative maps of all six lattice parameters and orientation of the unit cell for an epitaxial thin film. We present a general formalism for an unambiguous transfer of the SXDM data for at least three Bragg reflections into microscopic maps of the total strain and the lattice rotation. We have used the technique to map strain and rotation in a patterned InGaN/InGaN double layer with different indium concentrations bonded to a handling wafer. The maps reveal variations of strain and orientation on the micrometer scale as well as partial relaxation that involves a buckling at the edge of the patterned mesa structures. 

We have discussed potential contributions to the obtained maps and quantified specifically the effect of a varying indium concentration as well as the strain and rotation fields due to edge and screw threading dislocations. This way, we extracted maps of indium content for both (i.e. top and bottom) layers of the structure and characterized the distribution of threading dislocations, although the individual dislocations are not resolved in the experiment. As pointed out in the Appendix \ref{appendix}, resolving individual dislocations may be achieved with an \xray{} beam spot size that is approximately 5 times smaller than in our experiment, which is nowadays available at dedicated synchrotron beamlines.

We find that the dislocations in the top layer are inherited from the bottom layer, so that regrowth with a higher indium concentration does not result in a nucleation of additional dislocations. The data also shows that V-pits lead to a reduced indium incorporation which becomes most visible close to the mesa edge. There, the indium content is drastically reduced despite the edge relaxation that leads to a reduced compressive tensile strain.

We have discussed the limitations of the technique. The range of variations in $\varepsilon_{xz}$ and $\varepsilon_{xz}$ provides an estimate of the limited experimental accuracy as well as the applicability of the plane stress condition $\sigma_{iz}=0\,(i=1,2,3)$ on the microscopic scale. Finally, our works paves the way for a more routine application SXDM to study the strain distribution in epitaxial layers, microstructures and devices with the potential to probe buried layers or samples in a complex environment.

\begin{acknowledgments}
We acknowledge the ESRF for beamtime at beamline ID01, the Institut de Recherche Technologique (IRT) NanoElectronique for supporting the use of the ESRF, and Soitec for the furniture of the InGaNOS wafer and for their support. VMK thanks Alexander Belov (Institute of Crystallography, Moscow) for useful discussions. CR thanks Michael Hanke (PDI) for estimating shear stresses based on a depth inhomogeneity of the In concentration.

\end{acknowledgments}

\appendix
\section{Probability distributions and correlation functions of the strain
and rotation components for dislocation arrays}
\label{appendix}

\begin{figure}
    \includegraphics[width=.435\columnwidth]{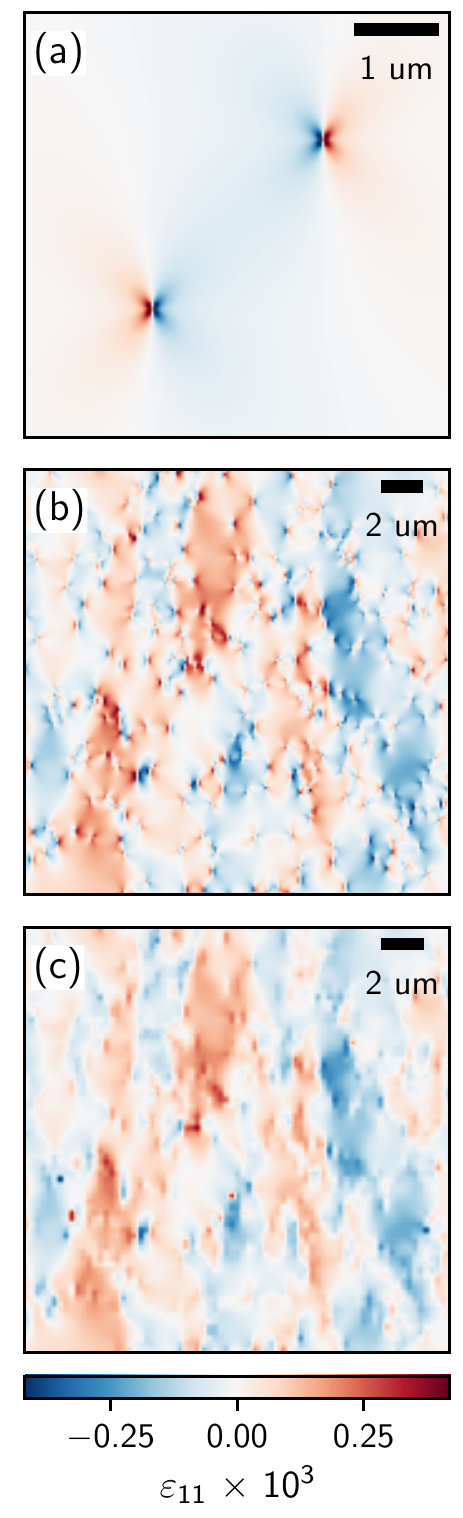}
    \includegraphics[width=.55\columnwidth]{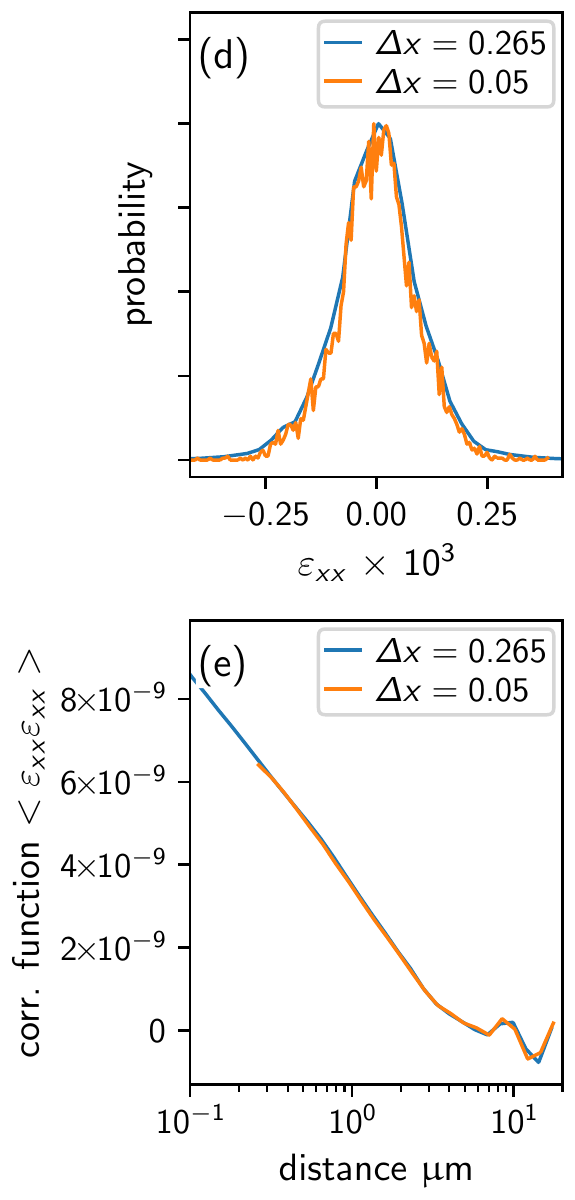}
    \caption{(a) Strain $\varepsilon_{xx}$ due to a pair of edge dislocations with the opposite Burgers vectors, (b) a $20\times20$~\textmu m$^{2}$ map of the strain $\varepsilon_{xx}$ due to such dislocation pairs randomly and uniformly distributed with the dislocation density $\varrho=1$~\textmu m$^{-2}$ and $M=8$, the pixel size is 0.05~\textmu m, (c) the same map for a pixel size of 0.265~\textmu m, (d) the strain probability densities and (e) the autocorrelation functions obtained from the two maps. The size of the maps is $20\times20 \unit{\upmu m ^2}$.}
\label{fig:MC-maps}
\end{figure}

In this Appendix, we describe the Monte Carlo modeling of the dislocation arrays. Fig.~\ref{fig:MC-maps}(a) shows the strain $\varepsilon_{xx}$ due to a dislocation pair consisting of two edge dislocations with the opposite Burgers vectors. Here and below, all calculations are made for a $2l=300$~nm thick free standing film, and the strain components are averaged over the interval $0<z<l$. Fig.~\ref{fig:MC-maps}(b) presents the strain $\varepsilon_{xx}$ due to such pairs of edge dislocations uniformly distributed with the dislocation density $\varrho=1$~\textmu m$^{-2}$ and $M=8$. The direction of the vector between the two dislocations of a pair is random, the Burgers vectors possess one of three orientations 120$^{\circ}$ with respect to each other. Since $M$ is larger than 1, the strain fields of the pairs overlap and the individual pairs cannot be recognized in the map. We take a lognormal distribution of the distances between dislocations in a pair, with the mean distance $R=M/\sqrt{\varrho}$. The standard deviation of this distribution is set to $R/2$.

Fig.~\ref{fig:MC-maps}(c) shows the same map as in Fig.~\ref{fig:MC-maps}(b) but with a 5 times worse resolution, obtained by averaging the strain over the pixel size of 0.265~\textmu m, representing the resolution of the experiment. The individual dislocations cannot be revealed anymore. However, the strain distribution is only little affected. Fig.~\ref{fig:MC-maps}(d) compares the strain probability distributions obtained from the maps in Figs.~ \ref{fig:MC-maps}(b) and \ref{fig:MC-maps}(c). The strain distribution is only smoothed out at a lower resolution of the map. Fig.~\ref{fig:MC-maps}(e) presents the autocorrelation function of the strain in the maps, considered below in detail. Here, we only note that the resolution has very little effect on the correlation function as well.

\begin{figure}
\includegraphics[width=1\columnwidth]{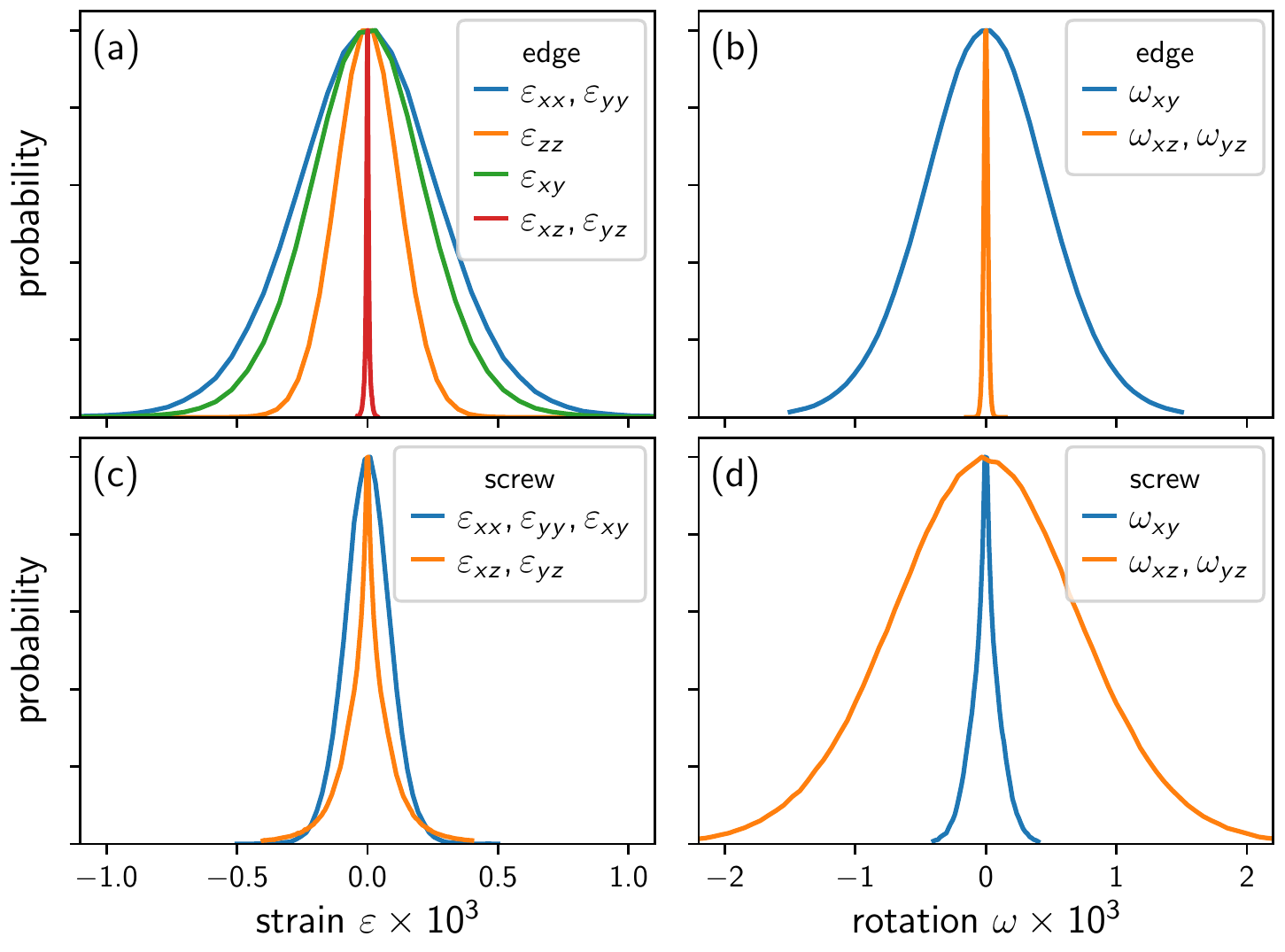}
\caption{Probability density distributions of the components of (a,c) strain and (b,d) rotation tensors for (a,b) edge and (c,d) screw dislocations. Dislocation density $\varrho=10$~\textmu m$^{-2}$ and $M=8$. Lateral sizes of the free standing film are $20\times20$~\textmu m$^{2}$ film, the strain and rotation components are averaged over a half $l=150$~\textmu m of the film thickness.}
\label{fig:histograms}
\end{figure}

\begin{figure}
\includegraphics[width=1.\columnwidth]{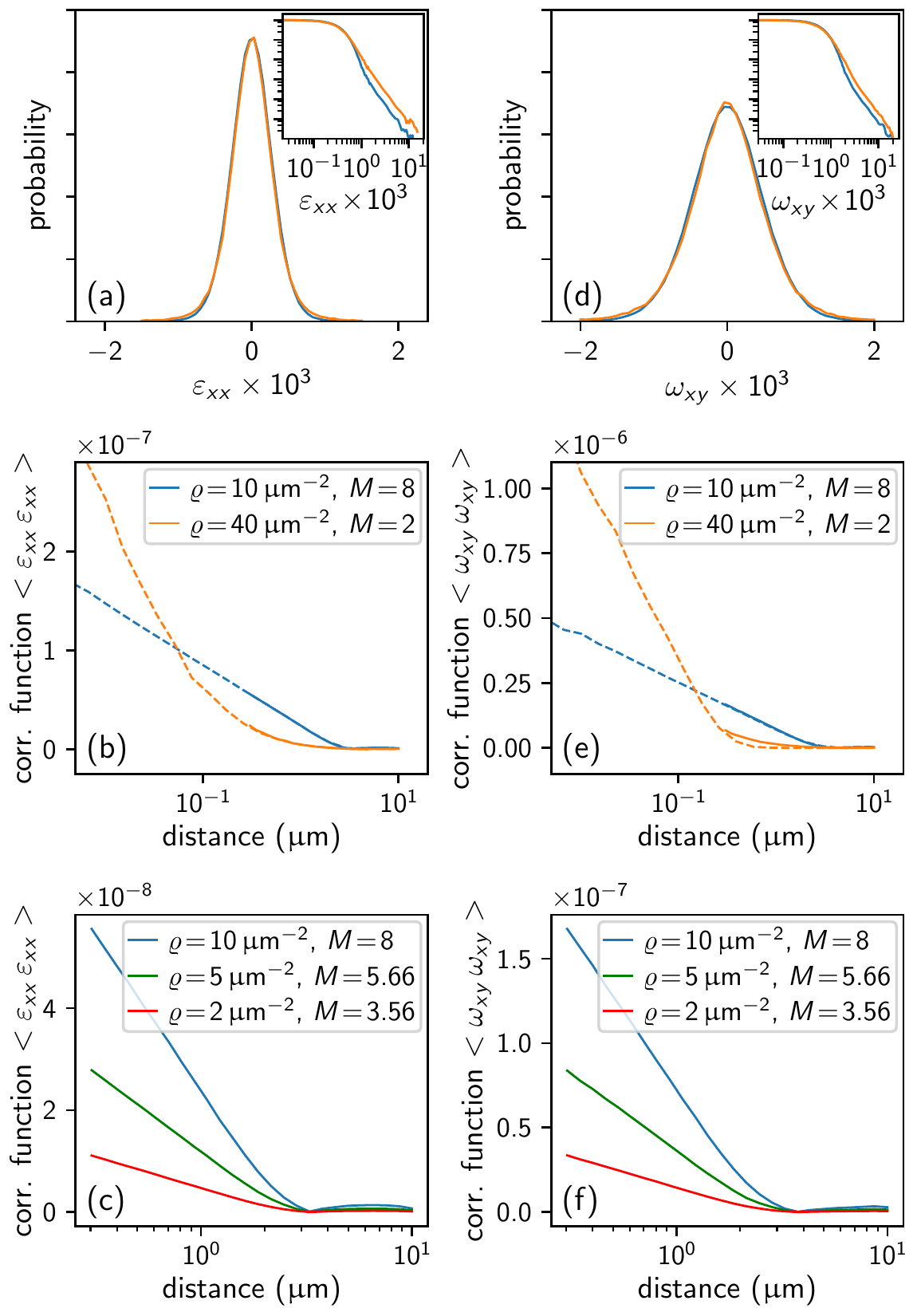}
\caption{(a) Probability density distributions of the strain $\varepsilon_{xx}$
for the dislocation arrays of edge dislocations with $\varrho=10$~\textmu m$^{-2}$, $M=8$ and $\varrho=40$~\textmu m$^{-2}$, $M=2$ (see legends in (b,e)). The inset shows the same distributions in the log-log scale. (b) The autocorrelation functions $\left\langle \varepsilon_{xx}\,\varepsilon_{xx}\right\rangle $ for the same dislocation arrays. The dashed line indicates the part of small distances that is not accessed in our experiment. (c) Autocorrelation functions $\left\langle \varepsilon_{xx}\,\varepsilon_{xx}\right\rangle $ for dislocation arrays with different dislocation densities $\varrho$ and the same radius $R=2.5$~\textmu m of the screening of the strain fields due to surrounding dislocations. (d) probability density distributions and (e,f) autocorrelation functions for the in-plane rotations $\omega_{xy}$ of the same dislocation arrays.}
\label{fig:autocorr}
\end{figure}

The dimensions of the maps presented in Figs.~\ref{fig:MC-maps}(b) and \ref{fig:MC-maps}(c) correspond to these of the experimental maps. As a result, the probability distribution in Fig.~\ref{fig:MC-maps}(d) and the autocorrelation function in Fig.~\ref{fig:MC-maps}(e) possess limited statistics. Monte Carlo modeling allows us to improve statistics by repeating the calculation many times and averaging the results. Such averaged quantities are presented in Figs.~\ref{fig:histograms} and \ref{fig:autocorr}.

Fig.~\ref{fig:histograms} shows the probability density distributions of all components of the strain and rotation tensors for edge and screw dislocations of the same density $\varrho=10$~\textmu m$^{-2}$ and $M=8$. Figs.~\ref{fig:histograms}(b) and \ref{fig:histograms}(d) show that the in-plane rotations $\omega_{xy}$ are almost entirely due to edge dislocations, while $\omega_{xz}$, $\omega_{yz}$ are due to screw dislocations. Hence, the analysis of these rotation components can be performed separately to obtain independently the densities of edge and screw dislocations.

Fig.~\ref{fig:autocorr}(a) presents the probability distributions of the strain $\varepsilon_{xx}$ for two arrays of edge dislocations differing in both dislocation densities and the distance of the dislocation pairs. These parameters are intentionally chosen so that the probability distributions practically coincide. The inset in Fig.~\ref{fig:autocorr}(a) shows the same probability distributions in the log-log scale, thus revealing the asymptotic of the probabilities at large strains. The curves shown in the inset requires substantial statistics and are obtained by repeating the calculation shown in Fig.~\ref{fig:MC-maps}(d) enough times for random arrays of dislocations. In the Stokes-Wilson approximation, these strain probability curves coincide with the intensity profiles measured by an ordinary \xray{} diffraction from the whole sample \cite{StokesWilson44,kaganer14acta}. In that case, the measurements provide sufficient dynamic range of intensities to obtain experimental curves similar to the ones presented in the inset in Fig.~\ref{fig:autocorr}(a). Fits of these curves then allow to determine both parameters, $\varrho$ and $M$ \cite{kaganer05GaN}. The probability density obtained in the experiment described in the present paper does not provide sufficient statistics to determine the two parameters of the dislocation ensemble unambiguously. This can, however, be achieved by additionally considering the correlation functions.


Fig.~\ref{fig:autocorr}(b) presents the autocorrelation functions
\begin{equation}
C(\mathbf{r})=\left\langle B(\mathbf{r}-\mathbf{r}')B(\mathbf{r}')\right\rangle ,\label{eq:53}
\end{equation}
where $B(\mathbf{r})$ denotes any component of strain or rotation produced by the whole dislocation ensemble. The average $\left\langle \ldots\right\rangle $ is performed over random positions and orientations of the dislocations. We denote these correlation functions as $\left\langle B\,B\right\rangle $ for brevity. Particularly, Fig.~\ref{fig:autocorr}(b) presents the autocorrelation function $\left\langle \varepsilon_{xx}\,\varepsilon_{xx}\right\rangle $.
In our model of the dislocation array as independent pairs of dislocations with opposite Burgers vectors, the total strain or rotation can be written as a sum over dislocation pairs,  
\begin{equation}
    B(\mathbf{r})=\sum_{j}\beta(\mathbf{r}-\mathbf{r}_{j}),\label{eq:54}
\end{equation}
where $\beta(\mathbf{r})$ is a strain or rotation component due to a dislocation pair. Since the dislocation pairs are independent, the correlation function is  
\begin{equation}
    C(\mathbf{r})=\varrho\left\langle \beta(\mathbf{r}-\mathbf{r}')\beta(\mathbf{r}')\right\rangle \label{eq:55}
\end{equation}
where the average $\left\langle \ldots\right\rangle $ is performed over possible Burgers vectors, orientation of a pair, distance between dislocations in a pair, and position $\mathbf{r}'$ of the pair in the plane of the layer. All these averages are performed simultaneously in the Monte Carlo calculation of the correlation functions.

Fig.~\ref{fig:autocorr}(b) presents the autocorrelation function $\left\langle \varepsilon_{xx}\,\varepsilon_{xx}\right\rangle $ for the same dislocation arrays as in Fig.~\ref{fig:autocorr}(a). As a consequence of the $\varepsilon\propto\rho^{-1}$ dependence of the strain at the dislocation line, the autocorrelation function possesses a $\propto\ln x$ dependence at small $x$, and hence the linear-log scale is used. A notable difference in the correlation functions for two dislocation distributions, that give indistinguishable strain probability distributions in Fig.~\ref{fig:autocorr}(a), is evident.

The correlation function follows the $\propto\ln x$ dependence as long as the distance $x$ between the correlated points remains smaller than the radius $R$ of the screening of the dislocation strain field by surrounding dislocations. Fig.~\ref{fig:autocorr}(c) compares the autocorrelation functions $\left\langle \varepsilon_{xx}\,\varepsilon_{xx}\right\rangle $ for different dislocation densities $\varrho$ and different values of $M$, chosen so that the screening radius $R=M/\sqrt{\varrho}$ remains the same, $R=2.5$~\textmu m. The correlation functions possess a linear decrease in the logarithmic scale, with the slope proportional to $\varrho$, as long as $x<R$. At larger separations, the correlations are absent.

Two correlation functions in Fig.~\ref{fig:autocorr}(b) possess different radii $R$ of the screening of the dislocation strains. The dislocation density $\varrho=10$~\textmu m$^{-2}$ with $M=8$ gives $R=2.5$~\textmu m, while $\varrho=40$~\textmu m$^{-2}$ with $M=2$ gives notably smaller screening radius $R=0.32$~\textmu m. The minimum distance $x$ presented in the plot is limited by the resolution of the present experiment, and the linear part of the curve is not reached. The dashed lines in Fig.~\ref{fig:autocorr}(b) extend the calculation of the correlation functions to smaller $x$ and shows that the available range of $x$ may not reveal all features of the curve. Nevertheless, a clear distinction between two curves in Fig.~\ref{fig:autocorr}(b) shows that the two parameters of the dislocation ensemble, $\varrho$ and $M$, can be unambiguously determined from the correlation functions.

Figs.~\ref{fig:autocorr}(d)--\ref{fig:autocorr}(f) presents a similar calculation of the probability distributions and the autocorrelation functions of the in-plane rotations $\omega_{xy}$ for the same dislocation arrays. They show that the considerations above are applicable to all components of the strain and rotation tensors of edge and screw dislocations. Finally, combining the results in Figs.~\ref{fig:histograms} and \ref{fig:autocorr}, we conclude that the dislocation distribution can be fully characterized by fitting of the autocorrelation functions. Since the rotations $\omega_{xy}$ and $\omega_{xz}$, $\omega_{yz}$ are due to solely edge and screw dislocations, respectively, the autocorrelation functions $\left\langle \omega_{xy}\,\omega_{xy}\right\rangle $ and $\left\langle \omega_{xz}\,\omega_{xz}\right\rangle $ are of primary interest to characterize the dislocation ensemble.

Summarizing, we find that only edge threading dislocations provide the in-plane rotations (twist, $\omega_{xy}$) and the in-plane strain, while screw threading dislocations give rise to the out of plane rotations (tilt, $\omega_{xz}$ and $\omega_{yz}$). Hence, edge and screw dislocations (or edge and screw components of mixed dislocations) can be determined separately from the respective probability distributions and autocorrelation functions. The use of only probability distributions does not allow to determine the dislocation density $\varrho$ unambiguously, since the probability distributions depend on two parameters, the density $\varrho$ and the dislocation correlations $M$. Plotting the autocorrelation functions in the linear-log scale, we directly obtain the range of dislocation correlations $R=M/\sqrt \varrho$. Hence, a simultaneous fit of the probability distributions and the autocorrelation functions allows to determine the dislocation density unambiguously. The Monte Carlo modeling in the present Appendix is made with the statistics required to obtain smooth curves. However, Fig.\ \ref{fig:ProbAutocorr} above shows that the limited statistics of our experiment is sufficient for quantitative analysis of the dislocation densities and the dislocation correlations.

\bibliography{InGaN_sxdm}

\end{document}


\maketitle

\section{Strain and rotation fields of threading dislocations in a film}

The aim of this section is to derive explicit expressions
for all components of the strain and rotation tensors for an edge
or a screw dislocation crossing a thin plate along its normal, with
the account of the elastic strain relaxation on both surfaces of the
film. Since the plate studied experimentally consists of two layers
with different In concentrations and hence different lattice parameters,
we derive the strains and rotations averaged over a half of the total
thickness of the film. Let $B(x,y,z)$ represents any component of
strain or rotation and the film of the thickness $2l$ is at $-l<z<l$.
Then, we calculate the average 
\begin{equation}
\bar{B}(x,y)=\frac{1}{l}\int_{0}^{l}B\,dz.\label{eq:15}
\end{equation}

\subsection{Expressions in cylindrical coordinates }

Expressions for strain components in cylindrical coordinates and their
transformation to Cartesian coordinates are presented in many books,
but the rotations are not written. It is worth to obtain all components
consistently. Gradient of a vector is (see Ref.~\citenum{slaughter02:book},
Eq.~(2.5.17)) 
\begin{eqnarray}
\nabla\mathbf{u} & = & \frac{\partial u_{r}}{\partial r}\mathbf{\hat{e}}_{\rho}\mathbf{\hat{e}}_{\rho}+\frac{1}{\rho}\left(\frac{\partial u_{\rho}}{\partial\phi}-u_{\phi}\right)\mathbf{\hat{e}}_{\rho}\mathbf{\hat{e}}_{\phi}+\frac{\partial u_{\rho}}{\partial z}\mathbf{\hat{e}}_{\rho}\mathbf{\hat{e}}_{z}\nonumber \\
 & + & \frac{\partial u_{\phi}}{\partial\rho}\mathbf{\hat{e}}_{\phi}\mathbf{\hat{e}}_{\rho}+\frac{1}{\rho}\left(\frac{\partial u_{\phi}}{\partial\phi}+u_{\rho}\right)\mathbf{\hat{e}}_{\phi}\mathbf{\hat{e}}_{\phi}+\frac{\partial u_{\phi}}{\partial z}\mathbf{\hat{e}}_{\phi}\mathbf{\hat{e}}_{z}\nonumber \\
 & + & \frac{\partial u_{z}}{\partial\rho}\mathbf{\hat{e}}_{z}\mathbf{\hat{e}}_{\rho}+\frac{1}{\rho}\frac{\partial u_{z}}{\partial\phi}\mathbf{\hat{e}}_{z}\mathbf{\hat{e}}_{\phi}+\frac{\partial u_{z}}{\partial z}\mathbf{\hat{e}}_{z}\mathbf{\hat{e}}_{z}.\label{eq:1}
\end{eqnarray}
Then, the components of the strain tensor are 
\begin{eqnarray}
\varepsilon_{\rho\rho} & = & \frac{\partial u_{\rho}}{\partial\rho}\nonumber \\
\varepsilon_{\phi\phi} & = & \frac{1}{\rho}\left(\frac{\partial u_{\phi}}{\partial\phi}+u_{\rho}\right)\nonumber \\
\varepsilon_{zz} & = & \frac{\partial u_{z}}{\partial z}\nonumber \\
\varepsilon_{\rho\phi} & = & \frac{1}{2}\left(\frac{1}{\rho}\frac{\partial u_{\rho}}{\partial\phi}-\frac{u_{\phi}}{\rho}+\frac{\partial u_{\phi}}{\partial\rho}\right)\label{eq:2}\\
\varepsilon_{\rho z} & = & \frac{1}{2}\left(\frac{\partial u_{\rho}}{\partial z}+\frac{\partial u_{z}}{\partial\rho}\right)\nonumber \\
\varepsilon_{\phi z} & = & \frac{1}{2}\left(\frac{\partial u_{\phi}}{\partial z}+\frac{1}{\rho}\frac{\partial u_{z}}{\partial\phi}\right),\nonumber 
\end{eqnarray}
and the components of the rotations tensor are 
\begin{eqnarray}
\omega_{\rho\phi} & = & \frac{1}{2}\left(\frac{1}{\rho}\frac{\partial u_{\rho}}{\partial\phi}-\frac{u_{\phi}}{\rho}-\frac{\partial u_{\phi}}{\partial\rho}\right)\nonumber \\
\omega_{\rho z} & = & \frac{1}{2}\left(\frac{\partial u_{\rho}}{\partial z}-\frac{\partial u_{z}}{\partial\rho}\right)\label{eq:3}\\
\omega_{\phi z} & = & \frac{1}{2}\left(\frac{\partial u_{\phi}}{\partial z}-\frac{1}{\rho}\frac{\partial u_{z}}{\partial\phi}\right).\nonumber 
\end{eqnarray}

The transformation from cylindrical to Cartesian components is given
by the rotation matrix 
\begin{equation}
\hat{P}=\left(\begin{array}{ccc}
\cos\phi & \sin\phi & 0\\
-\sin\phi & \cos\phi & 0\\
0 & 0 & 1
\end{array}\right).\label{eq:4}
\end{equation}
The (symmetric) strain tensor is transformed as $\hat{\varepsilon}^{(x,y,z)}=\hat{P}^{T}\hat{\varepsilon}^{(\rho,\phi,z)}\hat{P}$
and has components 
\begin{eqnarray}
\varepsilon_{xx} & = & \varepsilon_{\rho\rho}\cos^{2}\phi+\varepsilon_{\phi\phi}\sin^{2}\phi-2\varepsilon_{\rho\phi}\sin\phi\cos\phi\nonumber \\
\varepsilon_{yy} & = & \varepsilon_{\rho\rho}\sin^{2}\phi+\varepsilon_{\phi\phi}\cos^{2}\phi+2\varepsilon_{\rho\phi}\sin\phi\cos\phi\nonumber \\
\varepsilon_{zz} & = & \varepsilon_{zz}\label{eq:5}\\
\varepsilon_{xy} & = & \left(\varepsilon_{\rho\rho}-\varepsilon_{\phi\phi}\right)\sin\phi\cos\phi+\varepsilon_{\rho\phi}\left(\cos^{2}\phi-\sin^{2}\phi\right)\nonumber \\
\varepsilon_{xz} & = & \varepsilon_{\rho z}\cos\phi-\varepsilon_{\phi z}\sin\phi\nonumber \\
\varepsilon_{yz} & = & \varepsilon_{\rho z}\sin\phi+\varepsilon_{\phi z}\cos\phi,\nonumber 
\end{eqnarray}
while the (antisymmetric) tensor of rotations $\hat{\omega}^{(x,y,z)}=\hat{P}^{T}\hat{\omega}^{(\rho,\phi,z)}\hat{P}$
has only off-diagonal components 
\begin{eqnarray}
\omega_{xy} & = & \omega_{\rho\phi}\nonumber \\
\omega_{xz} & = & \omega_{\rho z}\cos\phi-\omega_{\phi z}\sin\phi\label{eq:6}\\
\omega_{yz} & = & \omega_{\rho z}\sin\phi+\omega_{\phi z}\cos\phi.\nonumber 
\end{eqnarray}

\subsection{Some useful formulas}

Below we use the following formulas for Bessel functions $J_{n}(x)$:
\begin{equation}
J_{2}(x)=\frac{2}{x}J_{1}(x)-J_{0}(x)\label{eq:7}
\end{equation}
\begin{equation}
\int_{0}^{\infty}\frac{dx}{x}J_{1}(x)=1\label{eq:8}
\end{equation}
\begin{equation}
\int_{0}^{\infty}\frac{dx}{x}J_{2}(x)=\frac{1}{2}\label{eq:9}
\end{equation}
\begin{equation}
\frac{\partial J_{0}(x)}{\partial x}=-J_{1}(x)\label{eq:10}
\end{equation}
\begin{equation}
\frac{\partial J_{1}(x)}{\partial x}=J_{0}(x)-\frac{1}{x}J_{1}(x).\label{eq:11}
\end{equation}

\begin{equation}
\frac{\partial^{2}J_{1}(x)}{\partial x^{2}}+\frac{1}{x}\frac{\partial J_{1}(x)}{\partial x}-\frac{1}{x^{2}}=-J_{1}(x)\label{eq:12}
\end{equation}

\subsection{Elastic field of a screw dislocation in the plate}

The displacement field of a screw dislocation in a plate of thickness
$2l$ is \cite{eshelby51,belov92} 
\begin{eqnarray}
u_{\phi} & = & -\frac{\partial\Psi}{\partial\rho}\nonumber \\
u_{z} & = & \frac{b_{z}}{2\pi}\phi,\label{eq:17}
\end{eqnarray}
where the stress function is 
\begin{equation}
\Psi=-\frac{b_{z}}{2\pi}\int_{0}^{\infty}\frac{dk}{k^{2}}\frac{\sinh kz}{\cosh kl}J_{0}(k\rho).\label{eq:18}
\end{equation}
Hence, 
\begin{equation}
u_{\phi}=-\frac{b_{z}}{2\pi}\int_{0}^{\infty}\frac{dk}{k}\frac{\sinh kz}{\cosh kl}J_{1}(k\rho).\label{eq:19}
\end{equation}

The non-zero components of stain and rotation tensors are 
\begin{eqnarray}
\varepsilon_{\rho\phi} & = & \frac{b_{z}}{4\pi}\int_{0}^{\infty}dk\,\frac{\sinh kz}{\cosh kl}J_{2}(k\rho)\nonumber \\
\varepsilon_{\phi z} & = & \frac{b_{z}}{4\pi}\left[\frac{1}{\rho}-\int_{0}^{\infty}dk\,\frac{\cosh kz}{\cosh kl}J_{1}(k\rho)\right]\nonumber \\
\omega_{\rho\phi} & = & \frac{b_{z}}{4\pi}\int_{0}^{\infty}dk\,\frac{\sinh kz}{\cosh kl}J_{0}(k\rho)\label{eq:20}\\
\omega_{\phi z} & = & -\frac{b_{z}}{4\pi}\left[\frac{1}{\rho}+\int_{0}^{\infty}dk\,\frac{\cosh kz}{\cosh kl}J_{1}(k\rho)\right].\nonumber 
\end{eqnarray}

When calculating the averages (\ref{eq:15}), we transform the expressions
to provide fast convergence of the integrals, for their numerical
calculation: 
\begin{eqnarray}
\bar{\varepsilon}_{\rho\phi} & = & \frac{b_{z}}{4\pi l}\left[\frac{1}{2}-\int_{0}^{\infty}\frac{ds}{s\cosh s}J_{2}(s\rho/l)\right]\nonumber \\
\bar{\varepsilon}_{\phi z} & = & \frac{b_{z}}{4\pi l}\left[\frac{l}{\rho}-1-\int_{0}^{\infty}\frac{ds}{s}(\tanh s-1)J_{1}(s\rho/l)\right]\nonumber \\
\bar{\omega}_{\rho\phi} & = & \frac{b_{z}}{4\pi l}\int_{0}^{\infty}\frac{ds}{s}\frac{\cosh s-1}{\cosh s}J_{0}(s\rho/l)\label{eq:21}\\
\bar{\omega}_{\phi z} & = & -\frac{b_{z}}{4\pi l}\left[\frac{l}{\rho}+1+\int_{0}^{\infty}\frac{ds}{s}(\tanh s-1)J_{1}(s\rho/l)\right].\nonumber 
\end{eqnarray}

Eshelby and Stroh \cite{eshelby51} derived a series expansion of
$u_{\phi}$ for $\rho\gg l$. The first terms of the expansion are
\begin{equation}
u_{\phi}\approx-\frac{b_{z}}{2\pi}\frac{z}{\rho}+\frac{2b_{z}}{\pi^{2}}\sqrt{\frac{l}{\rho}}\exp\left(-\frac{\pi\rho}{2l}\right)\sin\frac{\pi z}{2l}.\label{eq:22}
\end{equation}
Then, the components of the strain and the rotation tensors averaged
over $z$ are 
\begin{eqnarray}
\bar{\varepsilon}_{\rho\phi} & = & \frac{b_{z}}{4\pi}\left(\frac{l}{\rho}\right)^{2}\nonumber \\
\bar{\varepsilon}_{\phi z} & = & \frac{b_{z}}{\pi^{2}l}\sqrt{\frac{l}{\rho}}\exp\left(-\frac{\pi\rho}{2l}\right)\nonumber \\
\bar{\omega}_{\rho\phi} & = & \frac{b_{z}}{\pi^{2}l}\sqrt{\frac{l}{\rho}}\exp\left(-\frac{\pi\rho}{2l}\right)\nonumber \\
\bar{\omega}_{\phi z} & = & -\frac{b_{z}}{2\pi\rho}\hspace{3cm}(\rho\gg l).\label{eq:24}
\end{eqnarray}
It follows from Eq.~(\ref{eq:24}) that all strain and rotation components,
except $\bar{\omega}_{\phi z}$, decay faster than $1/\rho$ at large
distances $\rho$. In other words, the long-range strain for a screw
dislocation in a film is absent, and the only long-range rotation
is $\bar{\omega}_{\phi z}$.

\subsection{Elastic field of an edge dislocation in the plate}

Displacement field of an edge dislocation in the infinite crystal
has components (see Ref.~\citenum{belov92}, p.415) 
\begin{eqnarray}
u_{x}^{\infty} & = & \frac{b_{x}}{2\pi}\left(\phi+\frac{1}{4(1-\nu)}\sin2\phi\right)\nonumber \\
u_{y}^{\infty} & = & -\frac{b_{x}}{4\pi(1-\nu)}\left[(1-2\nu)\ln\rho-\sin^{2}\phi\right]\nonumber \\
u_{z}^{\infty} & = & 0.\label{eq:25}
\end{eqnarray}
In polar coordinates, the displacements are 
\begin{eqnarray}
u_{\rho}^{\infty} & = & u_{x}^{\infty}\cos\phi+u_{y}^{\infty}\sin\phi\nonumber \\
u_{\phi}^{\infty} & = & -u_{x}^{\infty}\sin\phi+u_{y}^{\infty}\cos\phi,\label{eq:26}
\end{eqnarray}
and nonzero strain components calculated by Eq.~(2) are 
\begin{eqnarray}
\varepsilon_{\rho\rho}^{\infty}=\varepsilon_{\phi\phi}^{\infty} & = & -\frac{b_{x}}{4\pi}\frac{(1-2\nu)}{(1-\nu)}\frac{\sin\phi}{\rho}\nonumber \\
\varepsilon_{\rho\phi}^{\infty} & = & \frac{b_{x}}{4\pi(1-\nu)}\frac{\cos\phi}{\rho},\label{eq:27}
\end{eqnarray}
and the only nonzero rotation component is 
\begin{equation}
\omega_{\rho\phi}^{\infty}=\frac{b_{x}}{2\pi}\frac{\cos\phi}{\rho}.\label{eq:28}
\end{equation}

The relaxation elastic field is determined by the stress function
(see Ref.~\citenum{belov92}, Eq. (157)) 
\begin{eqnarray}
\Phi & = & \frac{\nu b_{x}}{\pi(1-\nu)}\sin\phi\int_{0}^{\infty}\frac{dk}{k^{3}}\frac{J_{1}(k\rho)}{(\sinh2kl+2kl)}\label{eq:29}\\
 & \times & \left[\sinh kz(kl\cosh kl+2\nu\sinh kl)-kz\sinh kl\cosh kz\right].\nonumber 
\end{eqnarray}
The displacement components are 
\begin{equation}
u_{\rho}^{r}=\frac{\partial^{2}\Phi}{\partial\rho\partial z},\,\,u_{\phi}^{r}=\frac{1}{\rho}\frac{\partial^{2}\Phi}{\partial\phi\partial z},\,\,u_{z}^{r}=-2(1-\nu)\varDelta\Phi+\frac{\partial^{2}\Phi}{\partial z^{2}}.\label{eq:30}
\end{equation}
The strain component $\varepsilon_{\rho\rho}^{r}=\partial^{3}\Phi/\partial^{2}\rho\partial z$,
and the average over thickness (\ref{eq:15}) is $\bar{\varepsilon}_{\rho\rho}^{r}=l^{-1}\partial^{2}\Phi|_{z=l}/\partial^{2}\rho$:
\begin{equation}
\bar{\varepsilon}_{\rho\rho}^{r} =  \frac{2\nu^{2}b_{x}}{\pi(1-\nu)l}\sin\phi\int_{0}^{\infty}\frac{ds}{s}\frac{\sinh^{2}s}{\sinh2s+2s}\left[\frac{2J_{1}(s\rho/l)}{(s\rho/l)^{2}}-J_{1}(s\rho/l)-\frac{J_{0}(s\rho/l)}{s\rho/l}\right].\label{eq:31}
\end{equation}
At $\rho\gg l$, the integral in Eq.~(\ref{eq:31}) tends to $-l/(8\rho)$,
so that the total strain is 
\begin{equation}
\bar{\varepsilon}_{\rho\rho}^{\infty}+\bar{\varepsilon}_{\rho\rho}^{r}\approx-\frac{b_{x}}{4\pi}(1-\nu)\frac{\sin\phi}{\rho}\,\,\,\,(\rho\gg l).\label{eq:32}
\end{equation}
This solution for the state of plane stress is obtained from the plain
strain solution (\ref{eq:27}) by replacement of the Poisson ratio
$\nu$ by $\nu/(1+\nu)$ (see Ref.~\citenum{belov92}, p.439).

In a similar calculation for $\bar{\varepsilon}_{\phi\phi}^{r}$ component,
we take into account that $\partial^{2}\Phi/\partial\phi^{2}=-\Phi$
and obtain 
\begin{equation}
\bar{\varepsilon}_{\phi\phi}^{r}=\frac{1}{\rho l}\left(\frac{\partial\Phi_{z=l}}{\partial\rho}-\frac{\Phi_{z=l}}{\rho}\right),\label{eq:33}
\end{equation}
so that 
\begin{equation}
\bar{\varepsilon}_{\phi\phi}^{r} = \frac{2\nu^{2}b_{x}}{\pi(1-\nu)}\frac{\sin\phi}{\rho}\int_{0}^{\infty}\frac{ds}{s^{2}}\frac{\sinh^{2}s}{\sinh2s+2s}\left[J_{0}(s\rho/l)-\frac{2J_{1}(s\rho/l)}{s\rho/l}\right].\label{eq:34}
\end{equation}
At $\rho\gg l$, the integral in Eq.~(\ref{eq:34}) tends to $-1/8$,
which gives the same result (\ref{eq:32}) for the sum $\bar{\varepsilon}_{\phi\phi}^{\infty}+\bar{\varepsilon}_{\phi\phi}^{r}$.

To calculate $\bar{\varepsilon}_{\rho\phi}$, we substitute (\ref{eq:30})
into (\ref{eq:2}) and integrate over $z$: 
\begin{equation}
\bar{\varepsilon}_{\rho\phi}^{r}=\frac{1}{\rho l}\left(\frac{\partial^{2}\Phi_{z=l}}{\partial\rho\partial\phi}-\frac{\partial\Phi_{z=l}}{\partial\rho}\right).\label{eq:35}
\end{equation}
Substituting (\ref{eq:29}), we get 
\begin{equation}
\bar{\varepsilon}_{\rho\phi}^{r} = \frac{2\nu^{2}b_{x}}{\pi(1-\nu)}\frac{\cos\phi}{\rho}\int_{0}^{\infty}\frac{ds}{s^{2}}\frac{\sinh^{2}s}{\sinh2s+2s}
\left[J_{0}(s\rho/l)-\frac{2J_{1}(s\rho/l)}{s\rho/l}\right].\label{eq:36}
\end{equation}
Comparing with (\ref{eq:34}), we find $\bar{\varepsilon}_{\rho\phi}^{r}=\bar{\varepsilon}_{\phi\phi}^{r}\cot\phi$.
At $\rho\gg l$, the integral in Eq.~(\ref{eq:36}) tends to $-1/8$,
which gives the total strain 
\begin{equation}
\bar{\varepsilon}_{\rho\phi}^{\infty}+\bar{\varepsilon}_{\rho\phi}^{r}\approx\frac{b_{x}}{4\pi}(1+\nu)\frac{\cos\phi}{\rho}\,\,\,\,(\rho\gg l).\label{eq:37}
\end{equation}
It can be obtained from $\bar{\varepsilon}_{\rho\phi}^{\infty}$ in
Eq.~(\ref{eq:27}) by substitution of $\nu$ with $\nu/(1+\nu)$,
in the same way as above.

To calculate $\varepsilon_{zz}$, we take into account that $\partial^{2}\Phi/\partial\phi^{2}=-\Phi$
and hence 
\begin{equation}
u_{z}=-2(1-\nu)\left(\frac{\partial^{2}}{\partial\rho^{2}}+\frac{1}{\rho}\frac{\partial}{\partial\rho}-\frac{1}{\rho^{2}}\right)\Phi-(1-2\nu)\frac{\partial^{2}\Phi}{\partial z^{2}}.\label{eq:38}
\end{equation}
Then, differentiating Eq.~(\ref{eq:29}) we find 
\begin{equation}
\left.\frac{\partial^{2}\Phi}{\partial z^{2}}\right|_{z=l}=-\frac{1-\nu}{\nu}\Phi|_{z=l},\label{eq:39}
\end{equation}
and the average strain is 
\begin{equation}
\bar{\varepsilon}_{zz}=\frac{2\nu b_{x}}{\pi l}\sin\phi\int_{0}^{\infty}\frac{ds}{s}J_{1}(s\rho/l)\frac{\sinh^{2}s}{\sinh2s+2s}.\label{eq:40}
\end{equation}
In the limit $\rho\gg l$, the integral in Eq.~(\ref{eq:40}) tends
to $l/(4\rho)$. Then, in this limit, Eqs.~(\ref{eq:32}) and (\ref{eq:40})
give 
\begin{equation}
(1-\nu)\bar{\varepsilon}_{zz}+\nu(\bar{\varepsilon}_{\rho\rho}+\bar{\varepsilon}_{\phi\phi})=0\,\,\,\,\,(\rho\gg l).\label{eq:41}
\end{equation}
The averaged stress $\bar{\sigma}_{zz}$ is proportional to the expression
(\ref{eq:41}), so that the normal stress averaged over thickness
is zero far from the dislocation line.

To calculate $\varepsilon_{\rho z}$, we obtain using Eq.~(\ref{eq:30})
\begin{equation}
\varepsilon_{\rho z}=\frac{\partial}{\partial\rho}\left[\nu\frac{\partial^{2}\Phi}{\partial z^{2}}-(1-\nu)\left(\frac{\partial^{2}\Phi}{\partial\rho^{2}}+\frac{1}{\rho}\frac{\partial\Phi}{\partial\rho}-\frac{\Phi}{\varrho^{2}}\right)\right].\label{eq:44}
\end{equation}
Averaging over thickness with the stress function (\ref{eq:29}) gives
\begin{equation}
\bar{\varepsilon}_{\rho z}=\frac{\nu b_{x}\sin\phi}{\pi(1-\nu)}\frac{\partial}{\partial\rho}\int_{0}^{\infty}\frac{ds}{s^{2}}J_{1}(s\rho/l)\frac{(\cosh s-1)(\sinh s-s)}{\sinh2s+2s}.\label{eq:45}
\end{equation}
Differentiating over $\rho$, we get finally 
\begin{equation}
\bar{\varepsilon}_{\rho z} = \frac{\nu b_{x}\sin\phi}{\pi(1-\nu)l}\int_{0}^{\infty}\frac{ds}{s}\left[J_{0}(s\rho/l)-\frac{1}{s\rho/l}J_{1}(s\rho/l)\right] \frac{(\cosh s-1)(\sinh s-s)}{\sinh2s+2s}.\label{eq:46}
\end{equation}

To calculate $\varepsilon_{\phi z}$, we obtain using Eq.~(\ref{eq:30})
\begin{equation}
\varepsilon_{\phi z}=\frac{1}{\rho}\frac{\partial}{\partial\phi}\left[\nu\frac{\partial^{2}\Phi}{\partial z^{2}}-(1-\nu)\left(\frac{\partial^{2}\Phi}{\partial\rho^{2}}+\frac{1}{\rho}\frac{\partial\Phi}{\partial\rho}-\frac{\Phi}{\varrho^{2}}\right)\right].\label{eq:47}
\end{equation}
Comparing this expression with Eq.~(\ref{eq:44}) and proceeding
to the average strain $\bar{\varepsilon}_{\phi z}$, we obtain instead
of Eq.~(\ref{eq:45}) 
\begin{equation}
\bar{\varepsilon}_{\phi z}=\frac{\nu b_{x}\cos\phi}{\pi(1-\nu)\rho}\int_{0}^{\infty}\frac{ds}{s^{2}}J_{1}(s\rho/l)\frac{(\cosh s-1)(\sinh s-s)}{\sinh2s+2s}.\label{eq:48}
\end{equation}

Calculation of the rotation $\omega_{\rho\phi}^{r}$, using Eqs.~(3)
and (\ref{eq:30}) gives $\omega_{\rho\phi}^{r}=0$. Hence, $\omega_{\rho\phi}=\omega_{\rho\phi}^{\infty}$
given by Eq.~(\ref{eq:28}).

Calculation of the rotation $\omega_{\rho z}$ (since $\omega_{\rho z}^{\infty}=0$,
we have $\omega_{\rho z}=\omega_{\rho z}^{r}$) gives 
\begin{equation}
\omega_{\rho z}=-(1-\nu)\frac{\partial}{\partial\rho}\left[\frac{\partial^{2}\Phi}{\partial\rho^{2}}+\frac{1}{\rho}\frac{\partial\Phi}{\partial\rho}-\frac{\Phi}{\varrho^{2}}+\frac{\partial^{2}\Phi}{\partial z^{2}}\right].\label{eq:49}
\end{equation}
Averaging over thickness gives finally 
\begin{eqnarray}
\bar{\omega}_{\rho z} & = & \frac{2\nu b_{x}\sin\phi}{\pi l}\int_{0}^{\infty}\frac{ds}{s}\left[J_{0}(s\rho/l)-\frac{1}{s\rho/l}J_{1}(s\rho/l)\right]\nonumber \\
 &  & \times\frac{(\cosh s-1)\sinh s}{\sinh2s+2s}.\label{eq:50}
\end{eqnarray}
Similarly, calculation of the rotation $\omega_{\phi z}$ gives 
\begin{equation}
\omega_{\phi z}=\frac{(1-\nu)}{\rho}\frac{\partial}{\partial\phi}\left[\frac{\partial^{2}\Phi}{\partial\rho^{2}}+\frac{1}{\rho}\frac{\partial\Phi}{\partial\rho}-\frac{\Phi}{\varrho^{2}}+\frac{\partial^{2}\Phi}{\partial z^{2}}\right],\label{eq:51}
\end{equation}
and the average over thickness is 
\begin{equation}
\bar{\omega}_{\phi z}=\frac{\nu b_{x}\cos\phi}{\pi\rho}\int_{0}^{\infty}\frac{ds}{s^{2}}J_{1}(s\rho/l)\frac{(\cosh s-1)\sinh s}{\sinh2s+2s}.\label{eq:52}
\end{equation}

The strain and rotation components $\bar{\varepsilon}_{\rho z}$,
$\bar{\varepsilon}_{\phi z}$, $\bar{\omega}_{\rho z}$, and $\bar{\omega}_{\phi z}$
are absent in the plane stress solution and decay faster than $1/\rho$
at $\rho\gg l$.

\bibliographystyle{unsrtnat}
\bibliography{surface}

\newpage

\section{Auto-correlation functions}
The histograms and autocorrelation functions of rotations, indium content and strains are shown in Figures~1 and 2 for both top and bottom InGaN layers. Based on these, the values in Table~I of the main text have been calculated.
\begin{figure}[ht]
    \centering
    \includegraphics[width=0.49\textwidth]{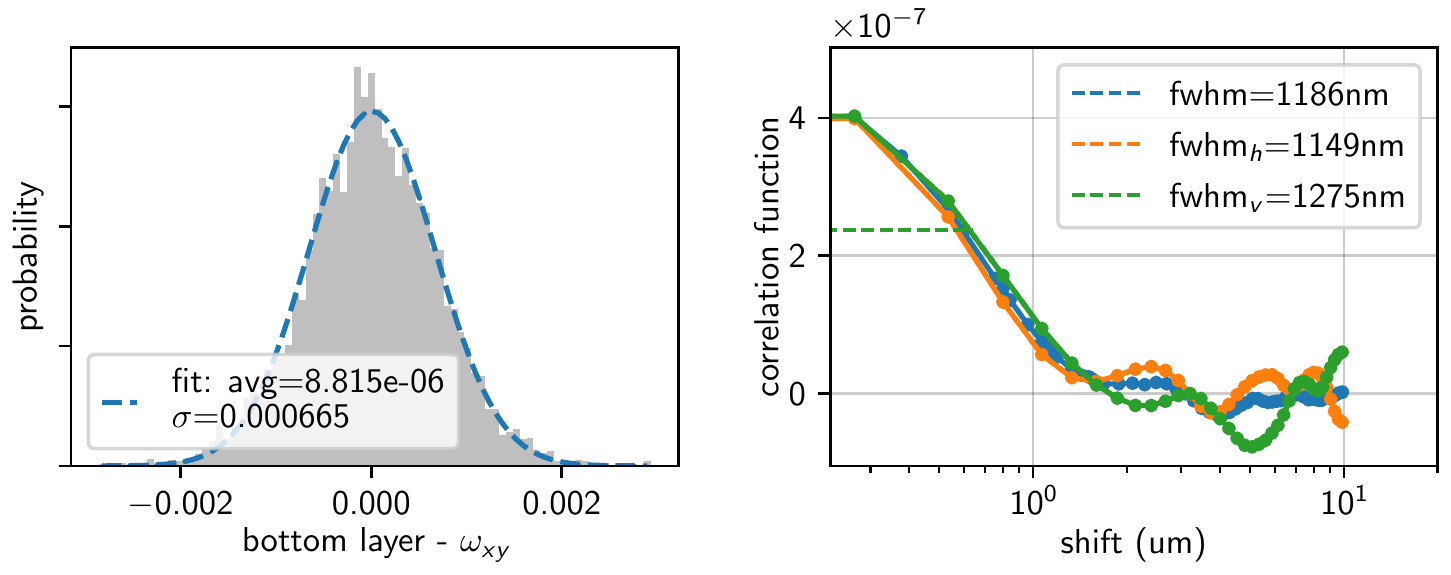}
    \includegraphics[width=0.49\textwidth]{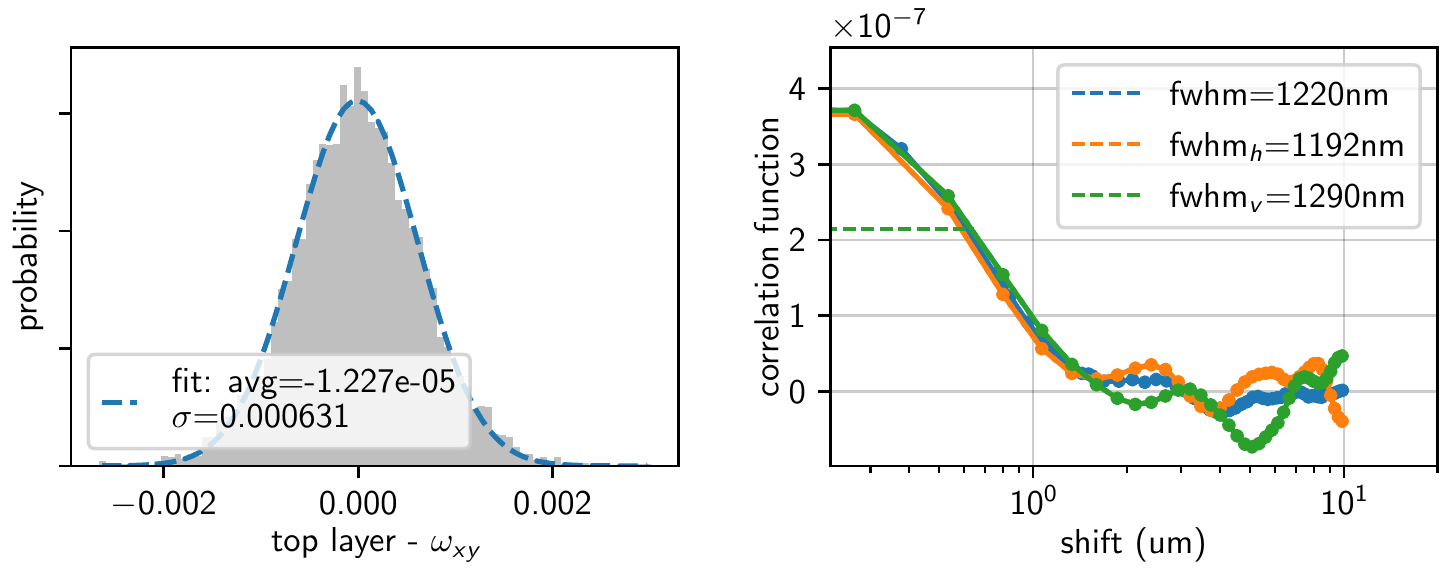}
    \includegraphics[width=0.49\textwidth]{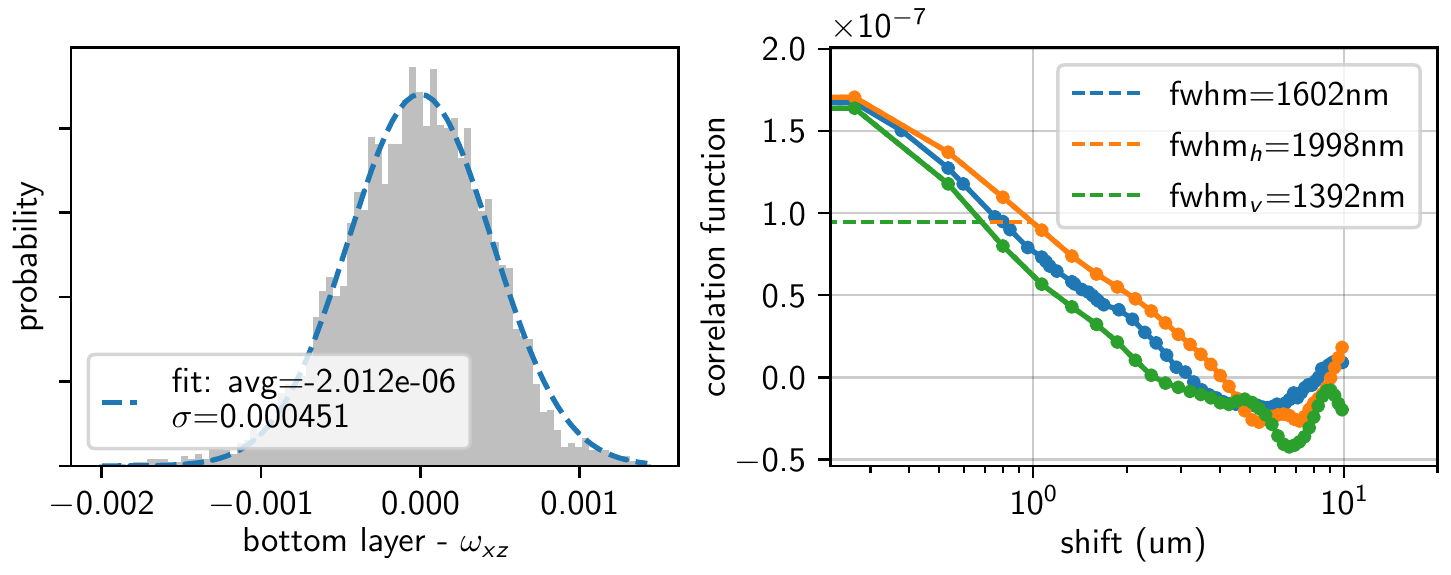}
    \includegraphics[width=0.49\textwidth]{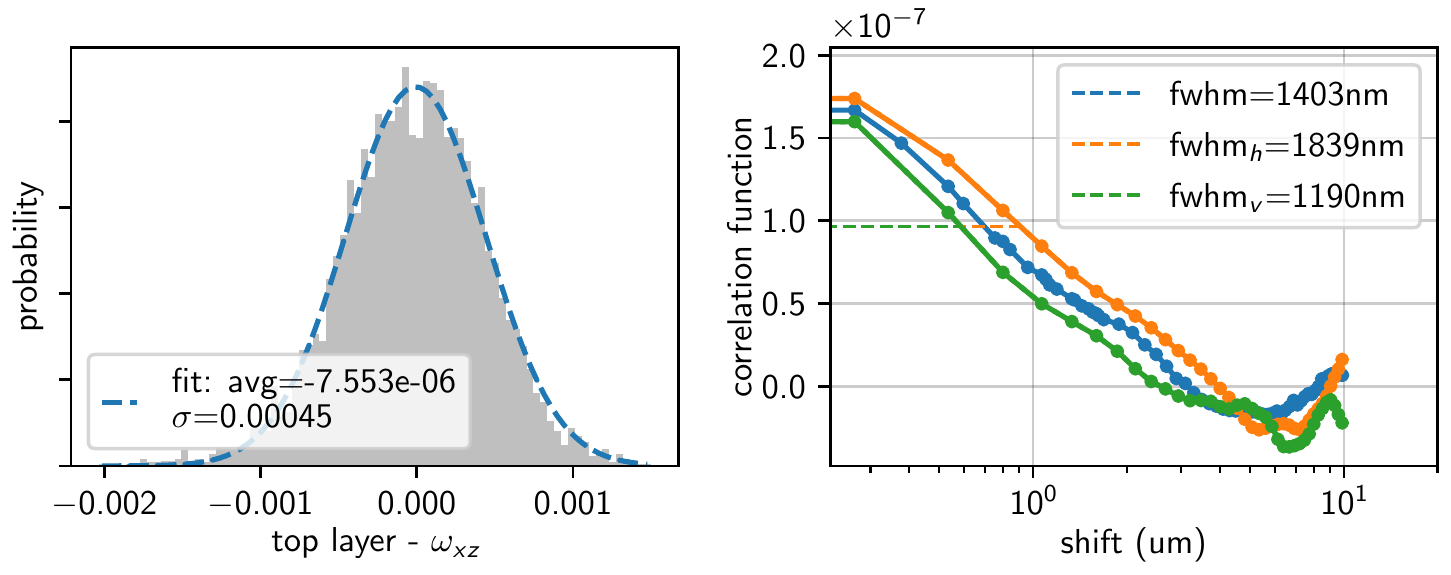}
    \includegraphics[width=0.49\textwidth]{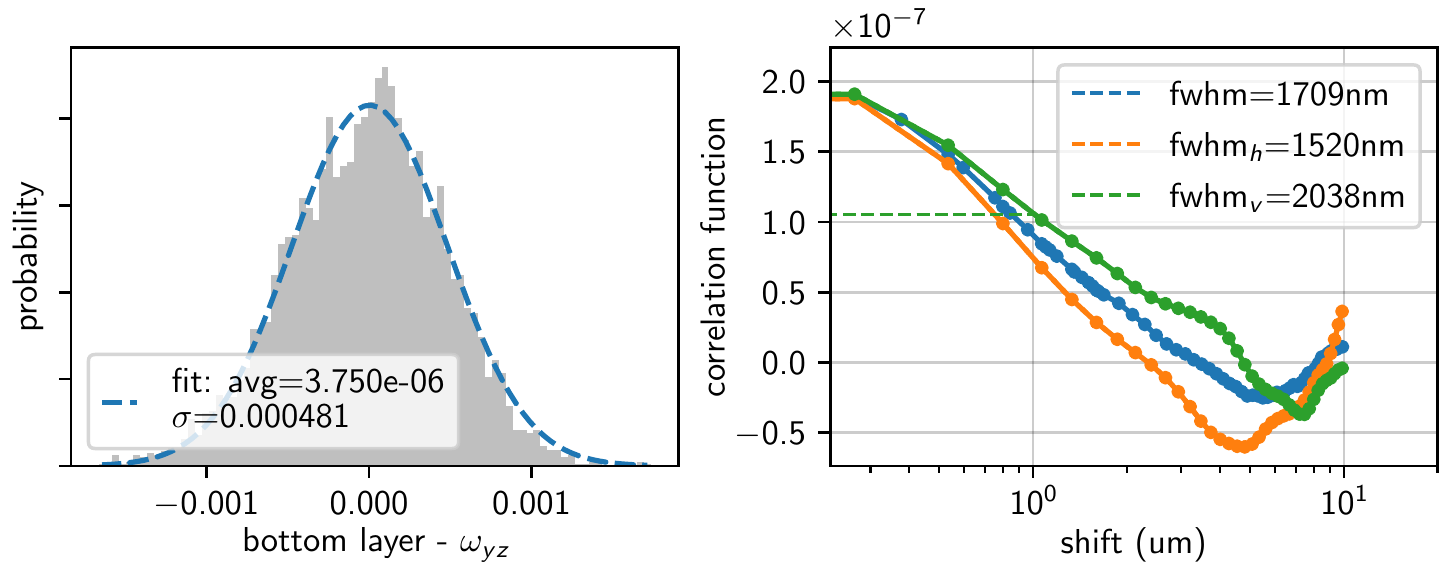}
    \includegraphics[width=0.49\textwidth]{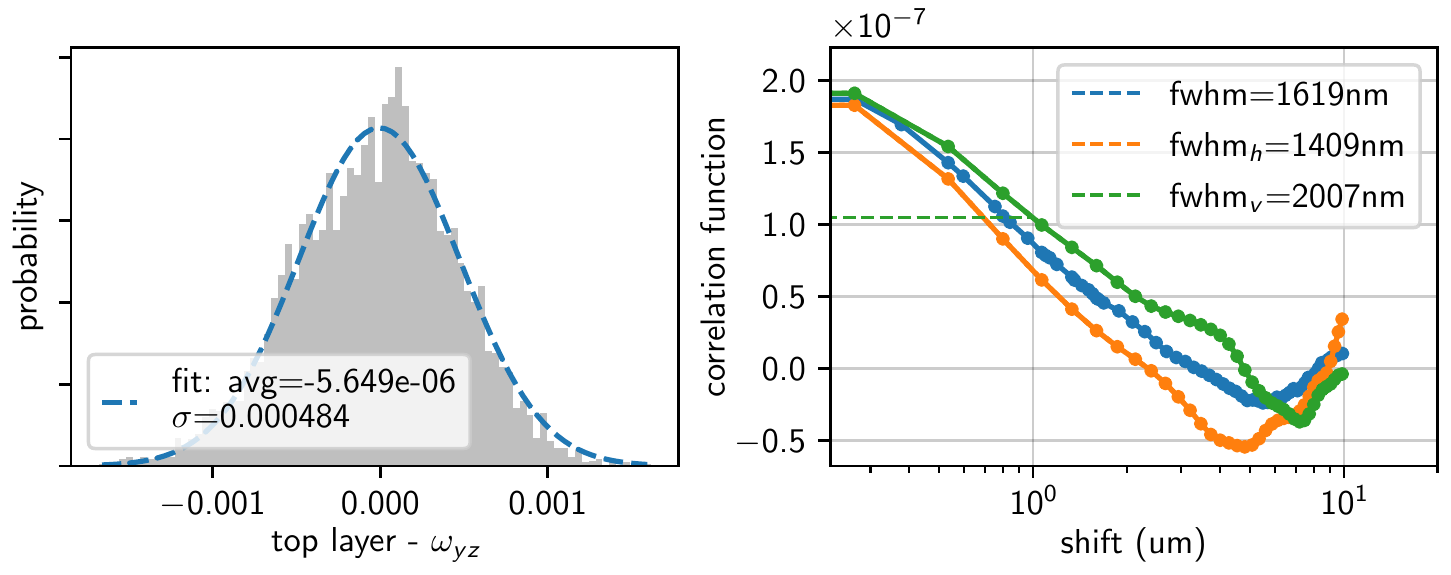}
    \includegraphics[width=0.49\textwidth]{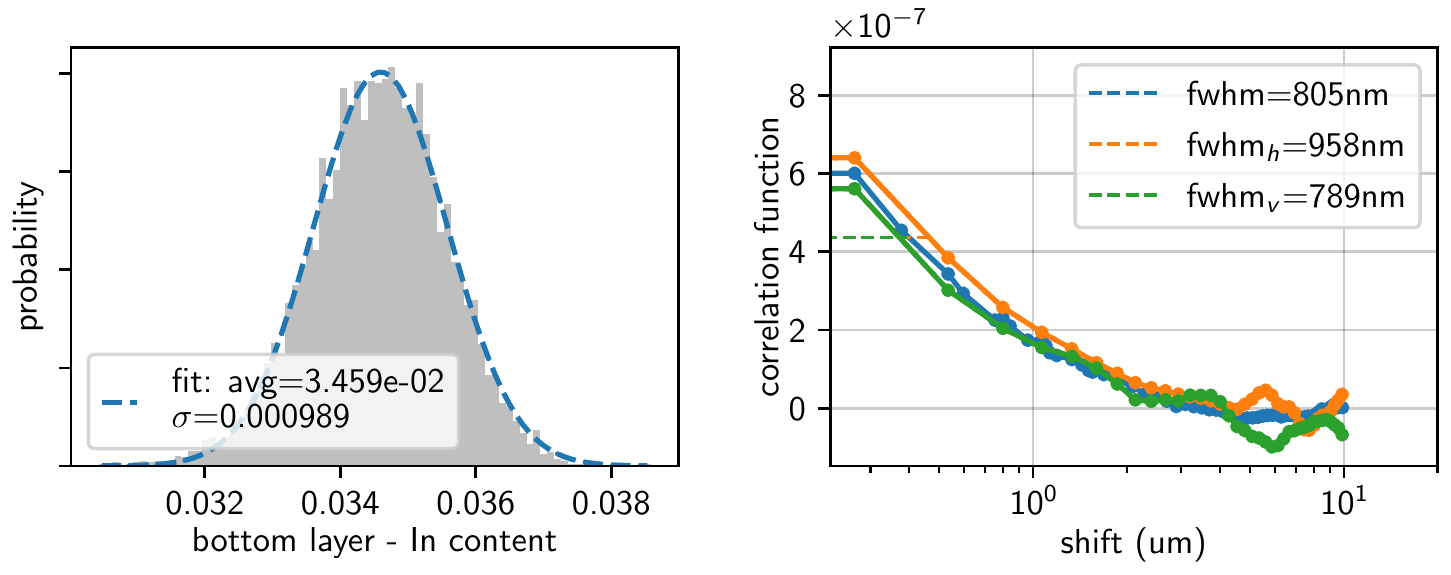}
    \includegraphics[width=0.49\textwidth]{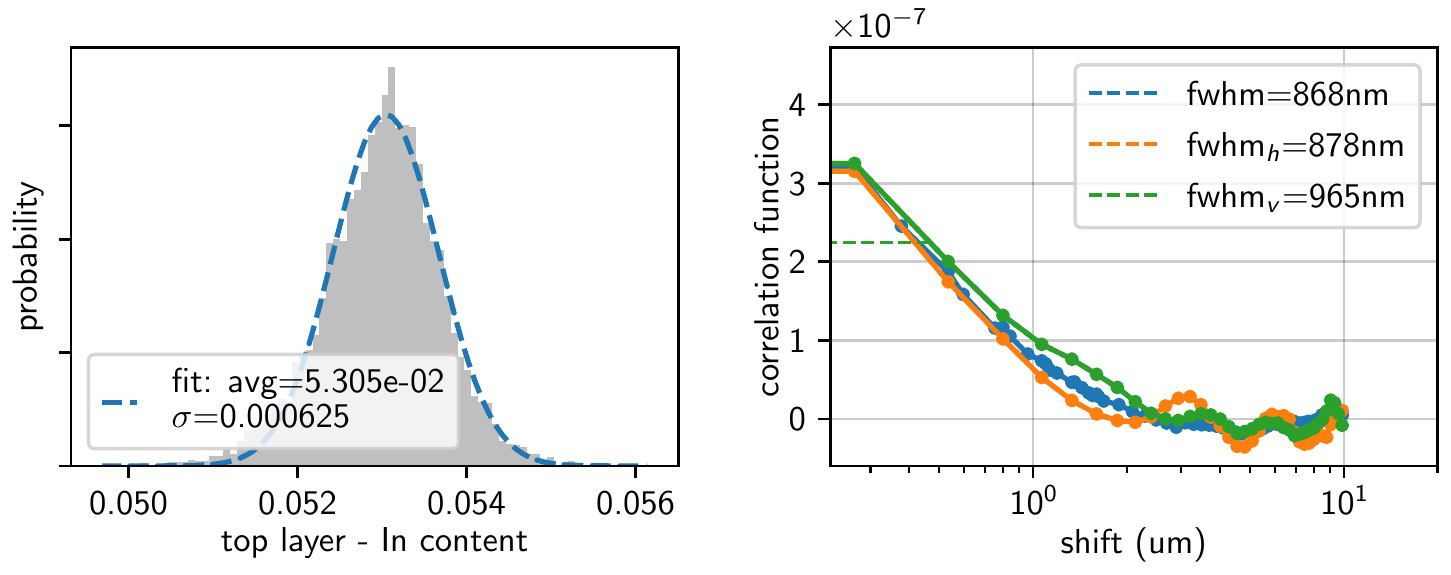}
    \caption{Histograms (column 1: bottom layer, column 3: top layer) and autocorrelation functions (column 2: bottom layer, column 4: top layer) of the rotation components (rows 1--3) and the indium content (row 4). Blue curves in the autocorrelation functions correspond to the radial dependence after azimuthal integration wheres orange and green curve correspond to the horizontal and vertical directions in the experimental maps, respectively. An anisotropy in the autocorrelation functions indicates an anisotropy in the corresponding rotation fields of dislocations. Legends in the histogram plots show the results of a Gaussian fit.
    }
    \label{fig:autocorrelation_rot}
\end{figure}

\begin{figure}[pht]
    \centering
    \includegraphics[width=0.49\textwidth]{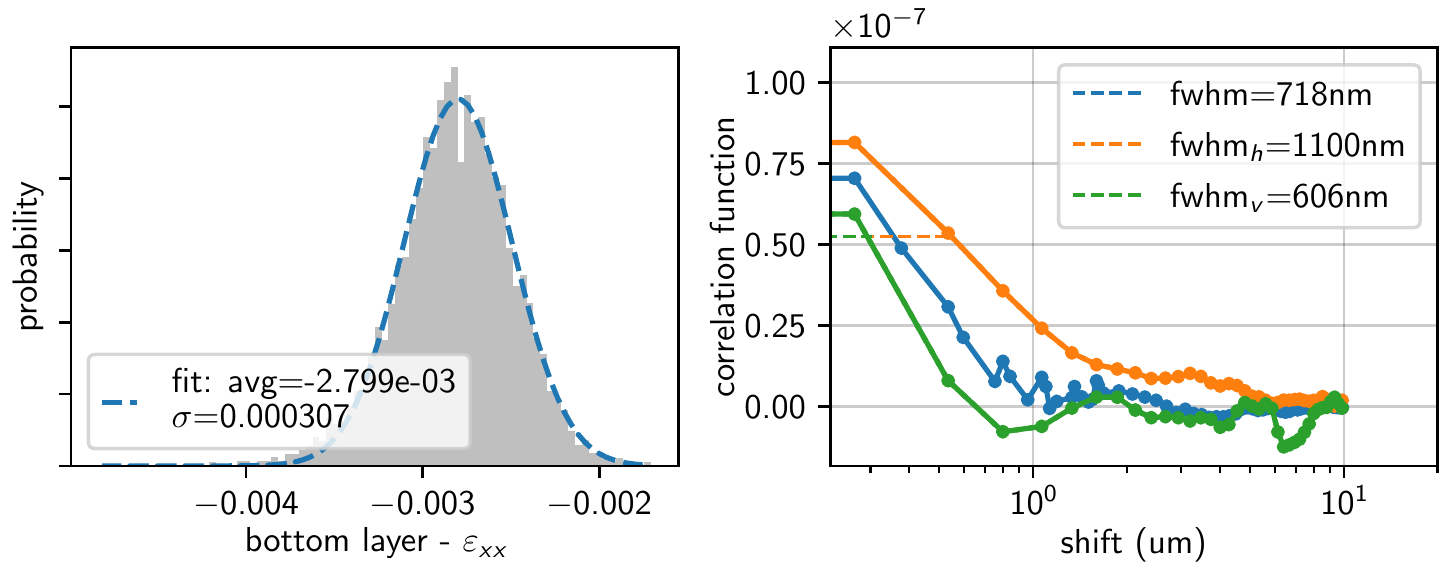}
    \includegraphics[width=0.49\textwidth]{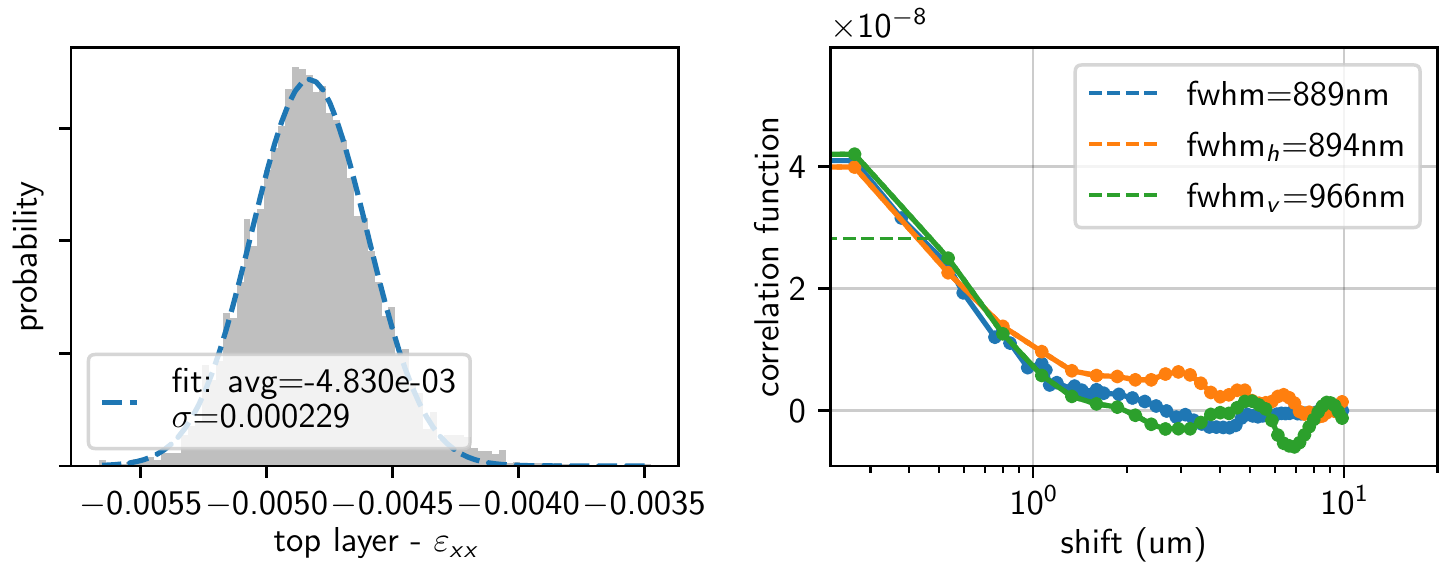}
    \includegraphics[width=0.49\textwidth]{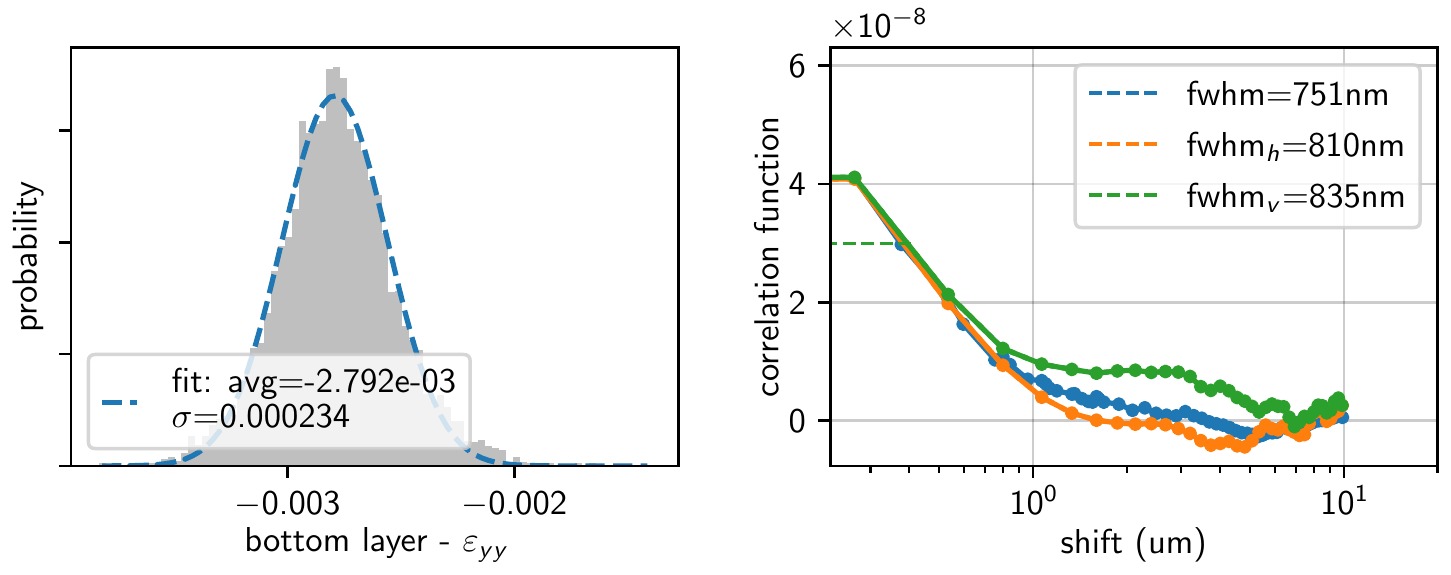}
    \includegraphics[width=0.49\textwidth]{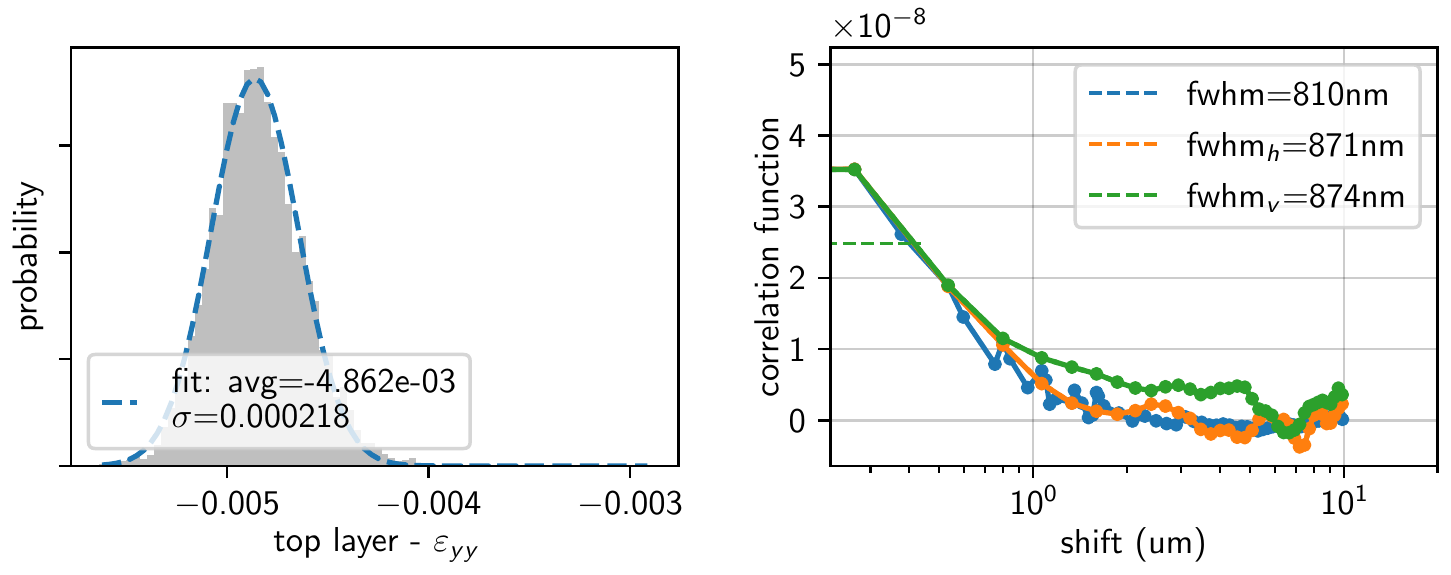}
    \includegraphics[width=0.49\textwidth]{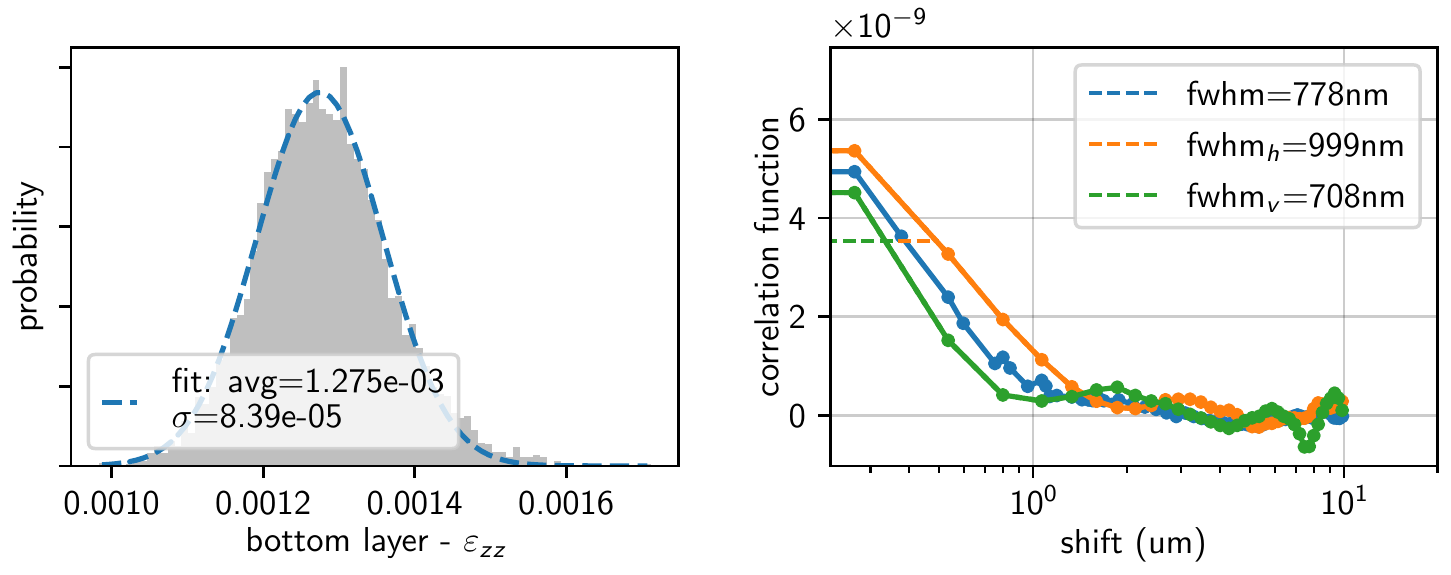}
    \includegraphics[width=0.49\textwidth]{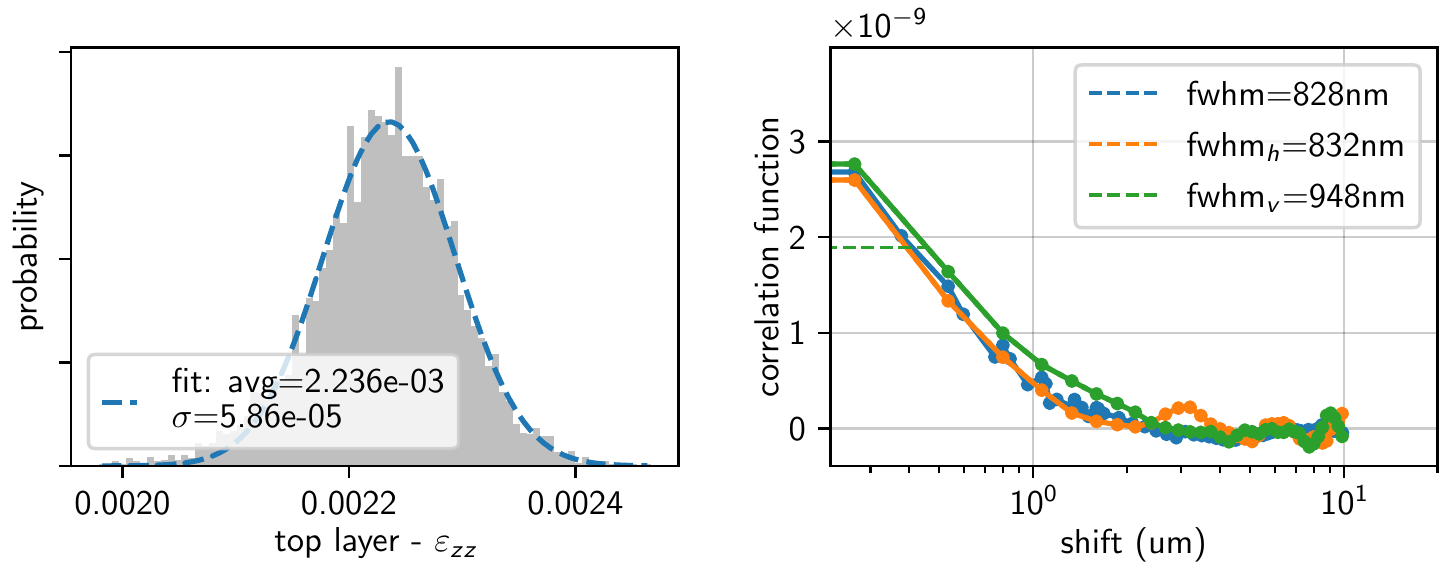}
    \includegraphics[width=0.49\textwidth]{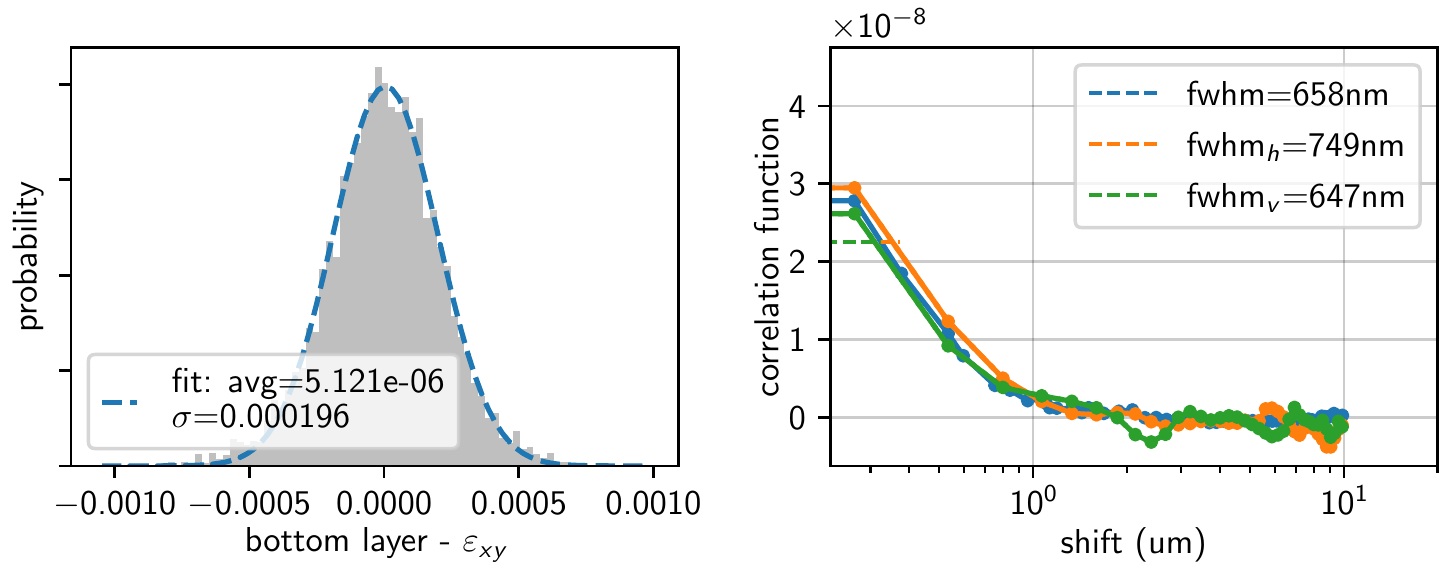}
    \includegraphics[width=0.49\textwidth]{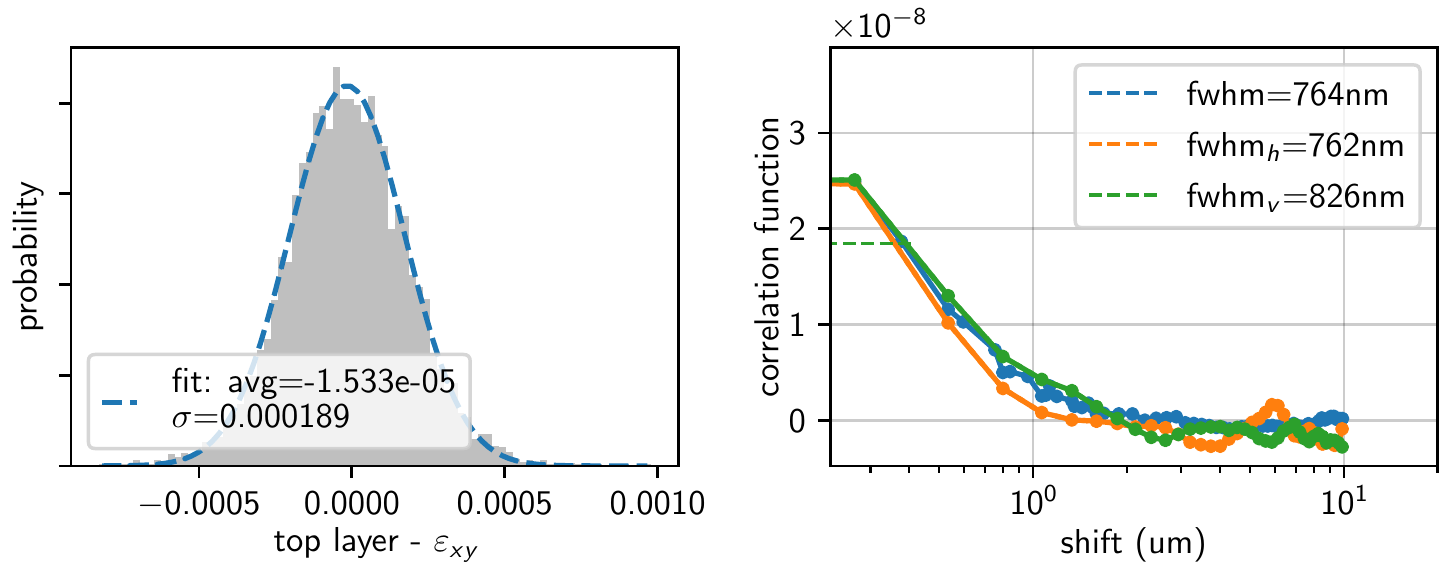}
    \includegraphics[width=0.49\textwidth]{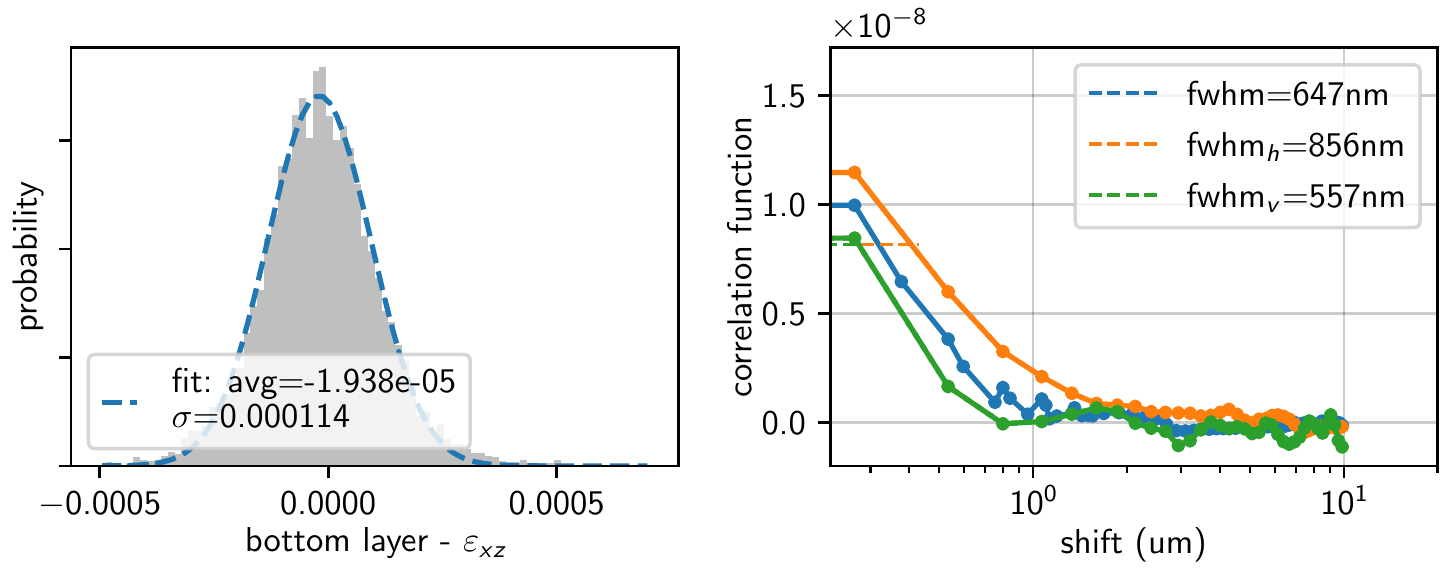}
    \includegraphics[width=0.49\textwidth]{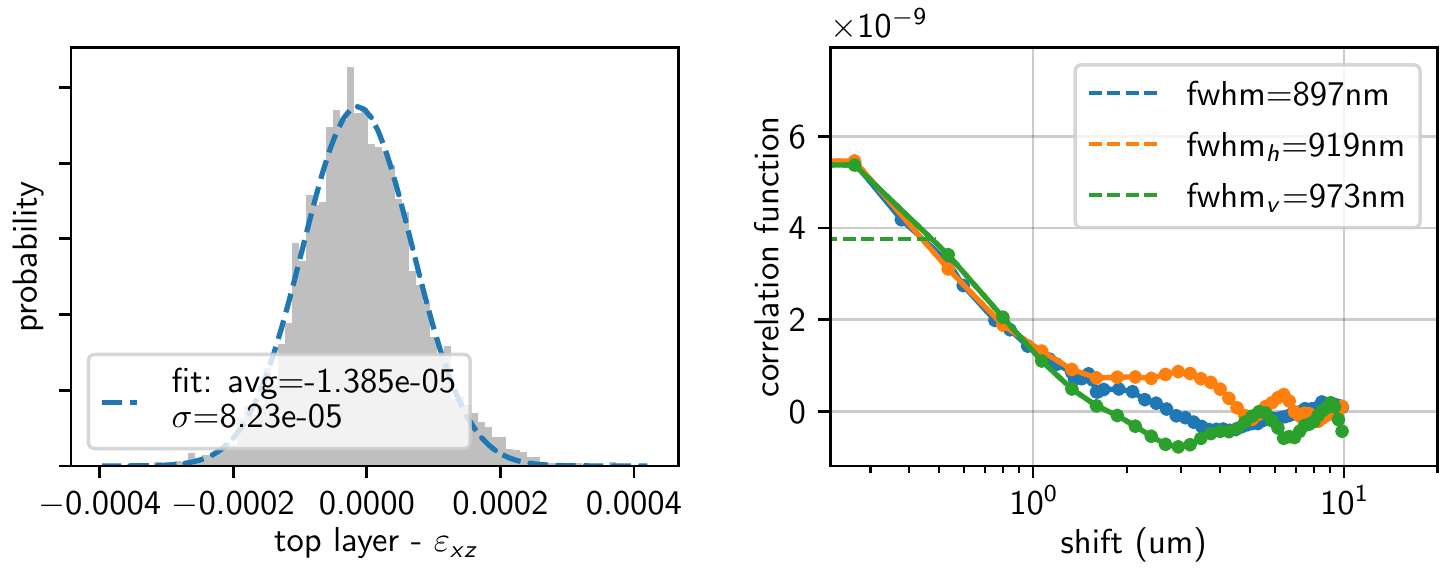}
    \includegraphics[width=0.49\textwidth]{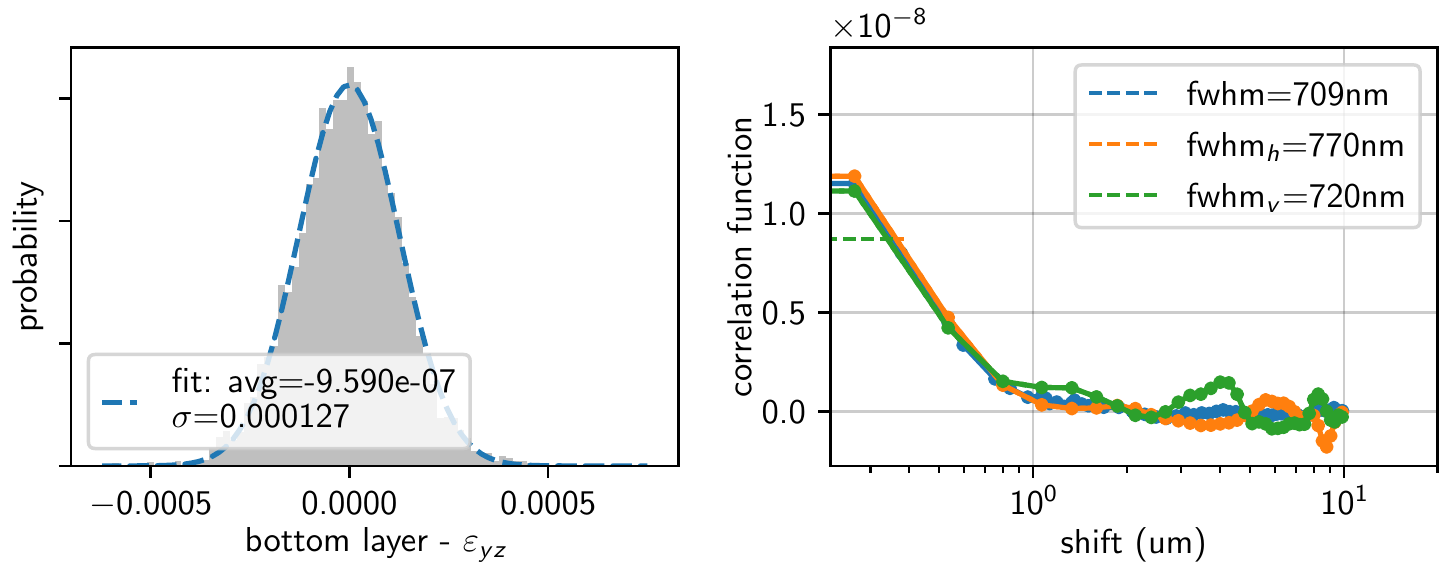}
    \includegraphics[width=0.49\textwidth]{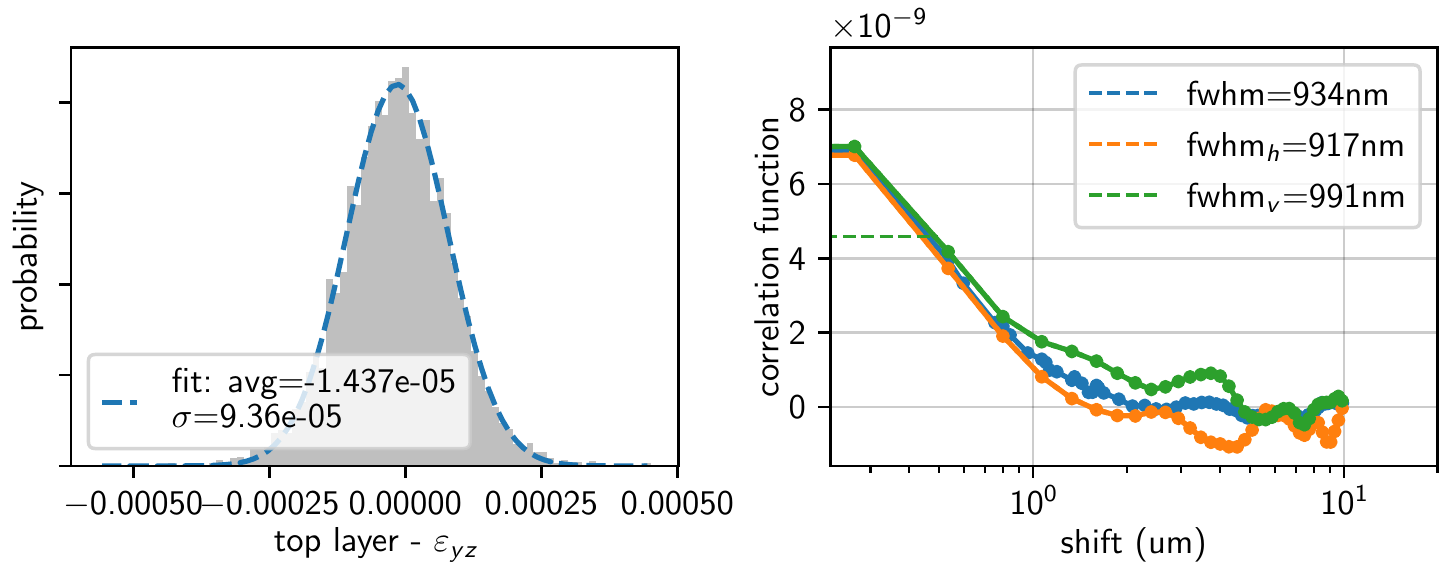}
    \caption{Histograms (column 1: bottom layer, column 3: top layer) and autocorrelation functions (column 2: bottom layer, column 4: top layer) of the six components of the elastic strain tensor. Blue curves in the autocorrelation functions correspond to the radial dependence after azimuthal integration, whereas orange and green curves correspond to the horizontal and vertical directions in the experimental maps, respectively. An anisotropy in the autocorrelation functions indicates an anisotropy in the corresponding strain fields of dislocations. Legends in the histogram plots show the results of a Gaussian fit.
    }
    \label{fig:autocorrelation_strain}
\end{figure}

\newpage

\section{Cross-correlation of rotation components}
Fig.~3 shows experimental cross-correlations between rotation components of top and bottom InGaN layers showing that the rotation fields and hence the distribution of dislocations is practically the same. The two-dimensional cross-correlation functions in the sub-figures~3(c) also show the symmetry of rotation fields from dislocations.
\begin{figure}[ht]
    \centering
    \includegraphics[width=1.\textwidth]{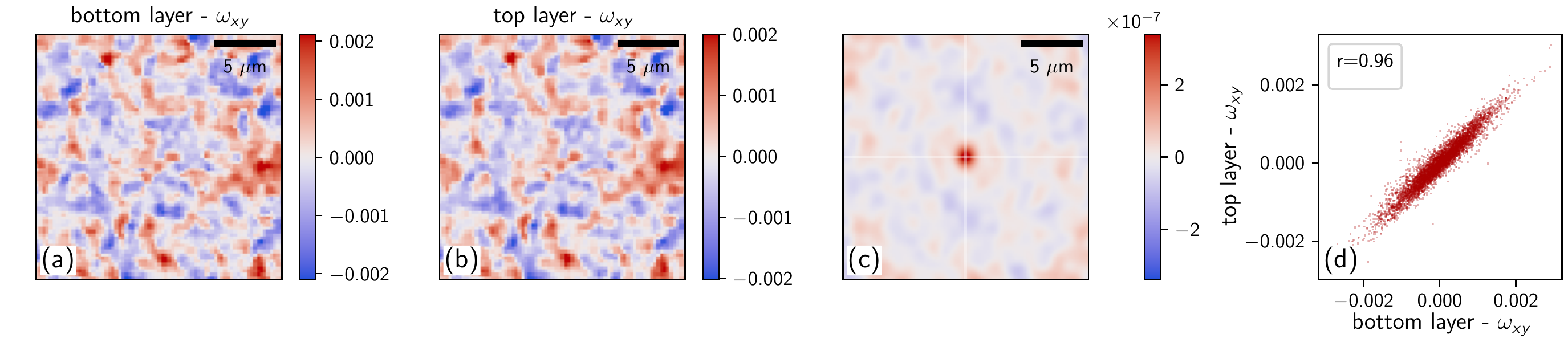}
    \includegraphics[width=\textwidth]{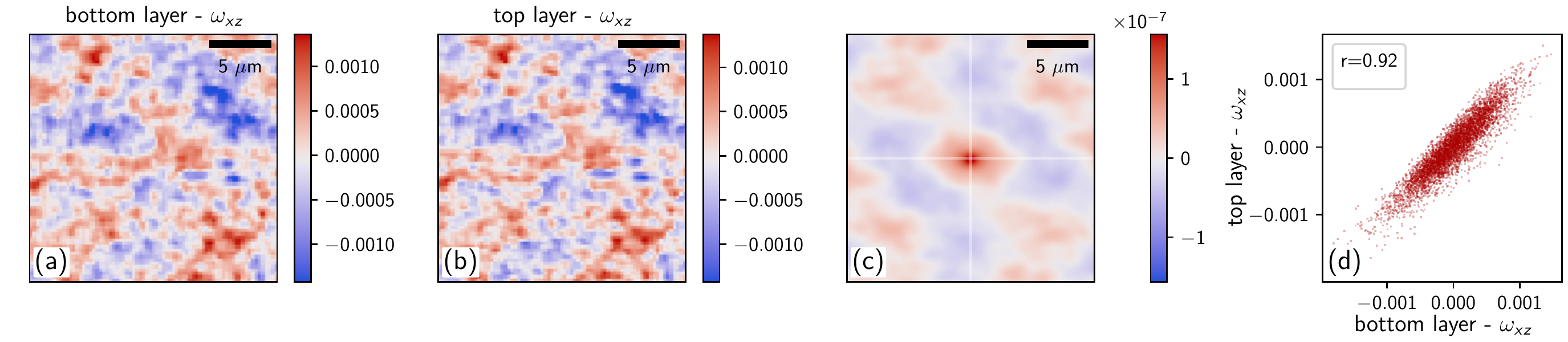}
    \includegraphics[width=\textwidth]{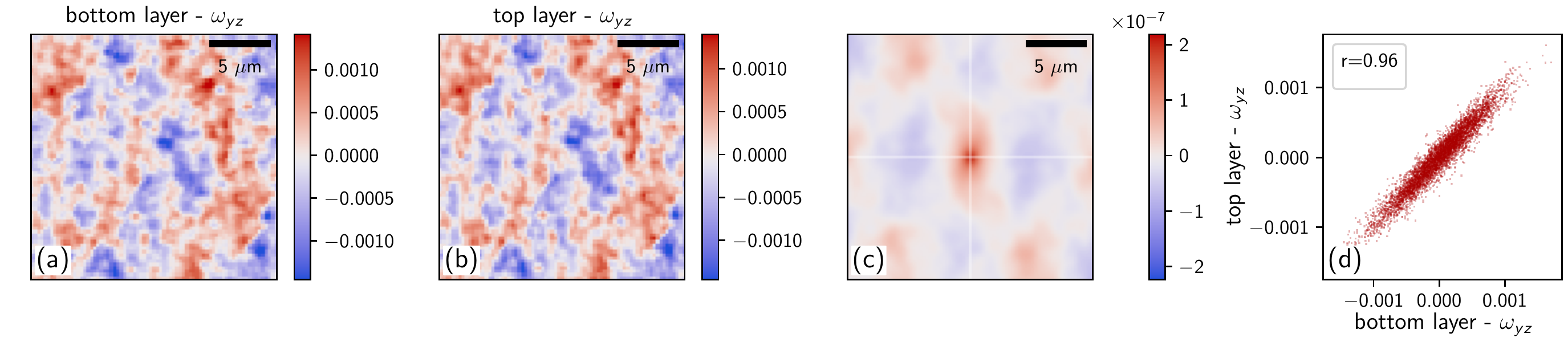}
    \caption{Comparison of the experimental maps of lattice rotations between the lower InGaN seed layer (column (a)) and the regrown top layer (column (b)). The two-dimensional correlation function is shown in column (c) and the relation of data points is shown in column (d) where the legend shows the obtained Pearson correlation coefficient $r$. The components $\omega_{xy}$, $\omega_{xz}$ and $\omega_{yz}$ are shown in the top, middle and bottom row, respectively. A high degree of similarity between seed layer and top layer is obvious from these figures indicating that the dislocation distribution is inherited from the seed during regrowth. The maps of rotation components give the clearest signature of the dislocation distribution since they are more pronounced compared to strain maps.}
    \label{fig:crosscorrelation_rot}
\end{figure}